\documentclass[aps,prd,nofootinbib,two column]{revtex4-2}
\usepackage{graphicx}
\usepackage{subcaption}
\usepackage{natbib}
\usepackage{mathtools}
\usepackage{slashed}
\usepackage{url}
\usepackage[normalem]{ulem}
\usepackage{enumitem}
\usepackage{mathrsfs}
\usepackage{float}
\usepackage{multirow}
\usepackage{xcolor}
\usepackage{booktabs}

\usepackage[colorlinks=true, linkcolor=blue, citecolor=violet, urlcolor=teal]{hyperref}
\newcommand{\orcid}[1]{\href{https://orcid.org/#1}{\includegraphics[width=8pt]
		{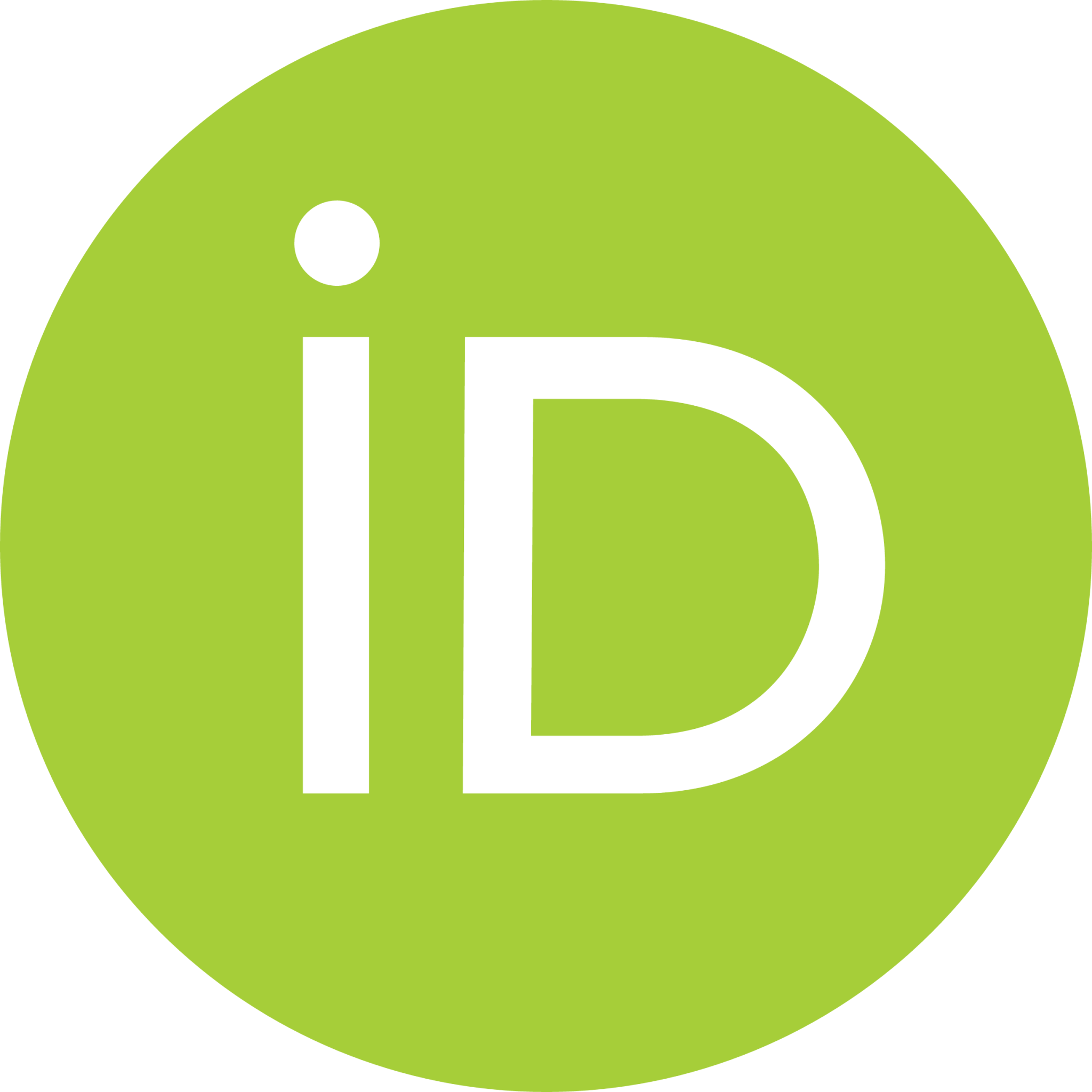}}}

\begin{document}

	\title{An expanding spherical fireball model for light hadron production at RHIC ($\sqrt{s_{\rm NN}}=7.7$--$39$ GeV)} 

\author{Ashutosh Dwibedi\orcid{https://orcid.org/0009-0004-1568-2806}}
\email{ashutoshdwibedi92@gmail.com}	
\affiliation{Department of Physics, Indian Institute of Technology Bhilai, Kutelabhata, Durg, 491002, Chhattisgarh, India}

\author{Anupam Panja\orcid{https://orcid.org/0009-0005-9692-0868}}
\email{anupampanja55@gmail.com}
\affiliation{Department of Physics, Indian Institute of Technology Bhilai, Kutelabhata, Durg, 491002, Chhattisgarh, India}

\author{Sabyasachi Ghosh\orcid{https://orcid.org/0000-0003-1212-824X}}
\email{sabya@iitbhilai.ac.in}
\affiliation{Department of Physics, Indian Institute of Technology Bhilai, Kutelabhata, Durg, 491002, Chhattisgarh, India}

\begin{abstract}
We investigate the transverse momentum ($p_T$) spectra and rapidity distributions of the light hadrons $\pi^{\pm}$, $K^{\pm}$, $p$, and $\bar{p}$ produced in Au+Au collisions at RHIC for $\sqrt{s_{\rm NN}} = 7.7$--39 GeV and different collision centralities. The produced medium is modeled as an expanding spherical fireball, with the radial expansion velocity determined from the rate of increase of the fireball radius. The particle spectra are calculated using the Cooper--Frye freeze-out prescription with a local equilibrium distribution function and a blast-wave-like flow profile. The model parameters are fixed from the midrapidity $p_T$ spectra of pions at kinetic freeze-out for different centralities. The same parameters are then used for the other hadron species, with the kinetic freeze-out chemical potential as the only additional free parameter. The model provides a good description of the STAR collaboration data for the $p_{T}$ spectra of light hadrons and predicts Gaussian-like rapidity distributions over the considered energy range across different centralities.

\end{abstract}

\maketitle
\section{Introduction}
Relativistic hydrodynamics has been remarkably successful in describing the space-time evolution of the matter produced in heavy-ion collisions (HICs) at Large Hadron Collider (LHC)~\cite{Roland:2014jsa,Foka:2016vta} and Relativistic Heavy Ion Collider (RHIC)~\cite{STAR:2017sal,Chen:2024aom}. The observed small shear viscosity-to-entropy density ratio and the strong collective flow indicate that the produced quark-gluon plasma (QGP) behaves as a nearly perfect fluid during the early stages of the collision~\cite{Shuryak:2003xe,Romatschke:2007mq}. Consequently, hydrodynamic models have successfully described the transverse momentum ($p_{T}$) spectra and anisotropic flow ($v_{n}$) over a broad range of collision energies and system sizes~\cite{Kolb:2003dz,Heinz:2013th,Gale:2013da,Romatschke:2017ejr,De:2022yxq,Ali:2024zvp,10.1093/ptep/pts014}.

Although relativistic hydrodynamics provides a realistic description of the bulk evolution, extracting freeze-out properties from full hydrodynamic simulations is computationally rich and requires detailed knowledge of the initial conditions and transport coefficients. As a result, hydrodynamics-inspired parameterizations, commonly known as blast-wave (BW) models, have been extensively employed to describe particle spectra and collective flow~\cite{PhysRevLett.42.880,Schnedermann:1993ws,Broniowski:2001we,Broniowski:2001uk,Huovinen:2001cy,Florkowski:2004tn,STAR:2001ksn,Retiere:2003kf,Tang:2008ud,He:2010vw,Sun:2014rda,Melo:2015wpa,Tomasik:2024uuq,Cimerman:2017lmm,Melo:2019mpn,FLORKOWSKI2025100249,Harabasz:2020sei,Harabasz:2022rdt,Drogosz:2025vdq,Teaney:2003kp,Jaiswal:2015saa,Yang:2016rnw,Yang:2018ghi,Yang:2020oig,Yang:2022ixy,Yang:2023apw,Rode:2018hlj,Rode:2020vhu,Chatterjee:2014lfa,Prasad:2021bdq,Panda:2025lmd,Parvan:2026idk,Alam:2026ixb}. Although the underlying philosophy is common to all BW models, they differ in their assumptions regarding the fluid flow profile, emission geometry, and freeze-out hypersurface. In the present work, we employ an expanding spherical fireball model to describe the $p_T$ spectra and rapidity distributions of identified hadrons produced in Au+Au collisions at RHIC for $\sqrt{s_{\rm NN}}=7.7$--39 GeV. The model is confronted with the data provided by STAR collaboration~\cite{STAR:2017sal} of $\pi^{\pm}$, $K^{\pm}$, $p$, and $\bar{p}$ over six collision centralities.

Fireball models have previously been employed to study electromagnetic and heavy-flavor probes~\cite{Rapp:1999us,Rapp:1999zw,Rapp:2000pe,Turbide:2003si,vanHees:2005wb,vanHees:2006ng,vanHees:2007th,vanHees:2011vb,Gossiaux:2011ea,vanHees:2014ida,Rapp:2014hha}, where an analytic description of the medium evolution is particularly advantageous. Motivated by these studies, we parameterize the fluid velocity of the medium with one important modification. Unlike the commonly adopted cylindrical geometry, we assume a spherically symmetric fireball with a radial velocity profile. The velocity at the fireball surface is identified with the expansion rate of the fireball radius (parameterized by $A$ and $v_{\infty}$), while the interior flow follows a BW-like parameterization. The idea of a spherically symmetric blast-wave model can be traced back to the pioneering works of Siemens and Rasmussen~\cite{PhysRevLett.42.880} and Bondorf \textit{et al.}~\cite{Bondorf:1978kz}. More recently, spherical and spheroidal geometries have also been successfully employed in the analysis of particle spectra and rapidity distribution at low collision energies~\cite{FLORKOWSKI2025100249,Harabasz:2020sei,Harabasz:2022rdt,Drogosz:2025vdq}.

 It is worth mentioning that in noncentral collisions, the pressure gradients along and perpendicular to the reaction plane are different, leading to anisotropic transverse expansion and consequently to anisotropic flow~\cite{Rai:2026dda}. However, the azimuthally integrated transverse momentum spectra and rapidity distributions are largely insensitive to these azimuthal anisotropies. Therefore, unlike the expanding elliptic fire-cylinder model of Ref.~\cite{Rai:2026dda}, which requires eight free parameters for the simultaneous description of spectra and anisotropic flow, the present model employs only five parameters. These are the chemical potential $\mu$ and temperature $T$ at kinetic freeze-out, two parameters ($A$, $v_{\infty}$) that characterize the time evolution of the radius of the expanding sphere and the freeze-out time $t_{f}$. With these parameters, we first fit the pion $p_{T}$ spectra at a given centrality to determine the common freeze-out parameters. The resulting freeze-out parameters are then kept fixed to describe the spectra of the remaining hadron species, allowing only the corresponding chemical potential to vary. Using the parameters extracted from the $p_{T}$ spectra, we subsequently calculate the rapidity distributions of all identified hadrons for each collision centrality. This strategy allows us to examine the extent to which a simple spherically expanding fireball can describe the transverse momentum spectra and reproduce the Gaussian-like rapidity distributions of identified hadrons across a broad range of RHIC Beam Energy Scan energies and collision centralities. It achieves this while employing fewer free parameters than more elaborate BW parametrizations~\cite{Rai:2026dda}.
 
The remainder of this paper is organized as follows. Sec.~\eqref{modeldesc} describes the model, Sec.~\eqref{results} presents the results, and Sec.~\eqref{summary} summarizes our findings. Throughout this paper, we use natural units $(c=\hbar=k_B=1)$ and adopt the mostly minus metric convention $(g_{\mu\nu}=(1,-1,-1,-1))$.

\section{Model Description}\label{modeldesc}
\begin{figure*}[t!]
	\centering
		\includegraphics[width=\linewidth]{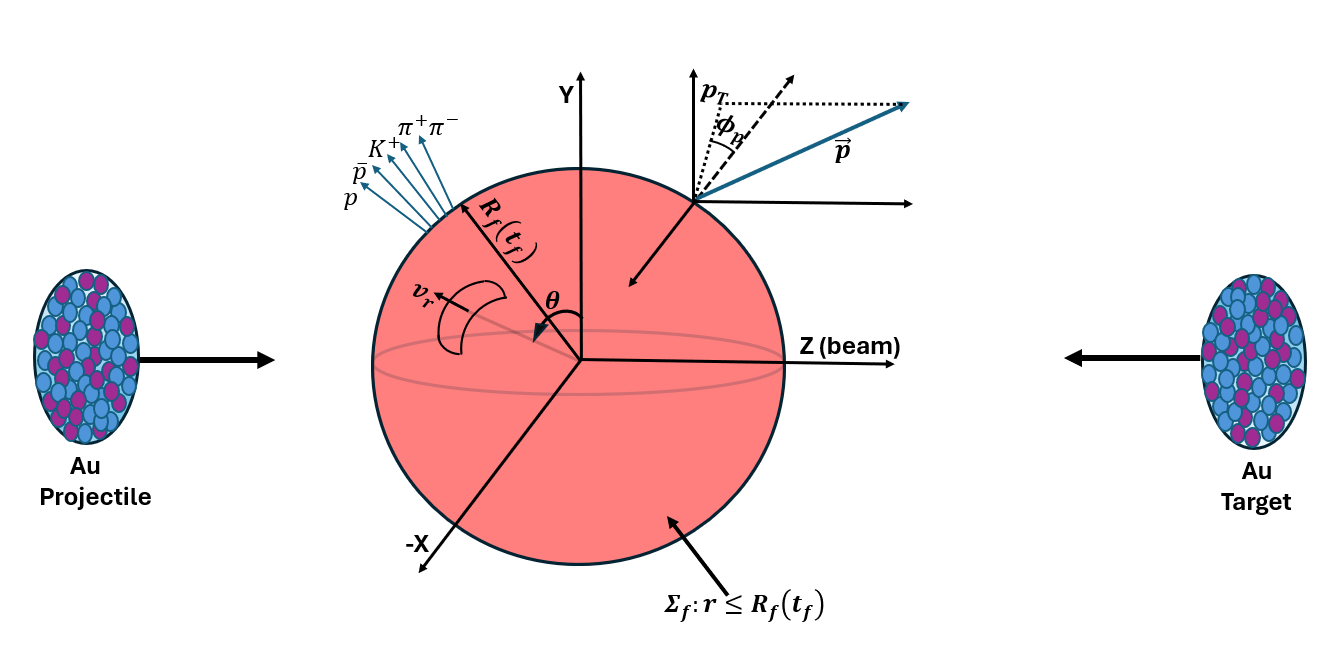}
\caption{Geometry of the expanding spherical fireball used in the present work. The freeze-out occurs on the spherical hypersurface $\Sigma_f$, defined by the boundary $r=R_f(t_f)$, where the fluid expands radially with velocity $v_r$. A fluid element on the freeze-out surface is specified by the polar angle $\theta$. The emitted particle momentum $\vec{p}$ is described by its transverse momentum $p_T$ and azimuthal angle $\phi_p$ with respect to the beam ($Z$) axis.}
	\label{fig:sphericalfireball}
\end{figure*}
Evaluating particle spectra emitted from the hot, dense medium in HIC requires modeling the space-time profile of the evolving fireball and prescribing the freeze-out hypersurface. We model the medium as an expanding spherical fireball with volume,
\begin{eqnarray}
	&&V(t)=\frac{4}{3}\pi r_{B}^{3}(t), \text{with increasing radius},\nonumber\\
	&&r_{B}(t)=r_{0}+v_{\infty}\bigg[t-\frac{1-e^{-At}}{A}\bigg]~.\label{A1}
\end{eqnarray}
The initial radius $r_{0}$ of the system can be determined from the collision specifications, i.e., by knowing the system of nuclei and the centrality class $(C)$. It is worth noting that the initial transverse dimension of the system for non-central collisions is elliptical, with a longitudinal spread determined by the energy of the colliding beams. This initial shape anisotropy evened out over time and turned into momentum anisotropy at the kinetic freeze-out, viz., the elliptic flow. In the present paper, since our primary aim is to explain the azimuthally integrated rapidity and transverse-momentum spectra, we do not pursue such shape anisotropies. However, the effective initial radius of the spherically symmetric fireball can be obtained from the initial elliptic cylinder as follows. The volume of the initial overlapping zone of the nuclei can be expressed as~\cite{Gossiaux:2011ea,Rai:2026dda} $V_{0}\equiv V(t=0)=2\pi~a_{0}b_{0}z_{0}$, which approximates the collision zone as a cylinder with elliptic cross-section $\pi~a_{0}b_{0}$. The initial transverse dimension of the system can be obtained from the geometry of two overlapping circles with radius $R$ in the $XY$ plane and given by,
\begin{eqnarray}
	&&b_{0}=R\left(1-\frac{C}{100}\right),\label{A2}\\
	&&a_{0}=\sqrt{2Rb_{0}-b_{0}^{2}}~.\label{A3}
\end{eqnarray}
The initial Lorentz contracted longitudinal size can be estimated as
\begin{eqnarray}
	&&z_{0}=\frac{a_{0}}{\gamma}=\frac{2a_{0}m_{N}}{\sqrt{s_{\rm NN}}},\label{A4}
\end{eqnarray}  
where $m_{N}$ and $\sqrt{s_{\rm NN}}$ are respectively, the mass of the nucleon and center of mass energy per nucleon pair. The effective initial radius of the spherical fireball is then duly expressed as,
\begin{eqnarray}
	&&r_{0}=\left(\frac{3V_{0}}{4\pi}\right)^{1/3}.\label{A5}
\end{eqnarray}
For Au$+$Au collisions, we take the radius $R=7$ fm.

The spherically symmetric fluid velocity of the system can be expressed as,
\begin{eqnarray}
	&&u^{\mu}=\cosh\eta_{r}(1,\tanh\eta_{r}\hat{r}),\label{A6}
\end{eqnarray}
where $\hat{r}=(\cos\theta~\hat{j}+\sin\theta \cos\phi~\hat{k}+\sin\theta \sin\phi ~\hat{i})$ is the unit vector along the radius and $\eta_{r}$ is the radial fluid rapidity. For convenience in the position space, we are measuring the polar angle $\theta$ about the $Y$ axis and the azimuthal angle $\phi$ in the reaction plane ($XZ$ plane) (see Fig.~\eqref{fig:sphericalfireball}). The fluid velocity (or the radial fluid rapidity) of the system can be specified by adhering to two simple aspects of the expansion dynamics, which fix the velocity at the surface and interior of the fireball. The surface velocity of the system is based on the rate of increase of the system radius in accordance with the spirit of the fireball model. The interior velocity can be designed by following the BW literature, which suggests an increasing radial velocity profile from the center to the surface. The radial fluid velocity can be written as,
\begin{eqnarray}
	&&v_{r}\equiv\tanh\eta_{r}=\frac{r}{r_{B}(t)} \dot{r}_{B}(t), \text{where}, \label{A7}\\
	&& \dot{r}_{B}(t)=v_{\infty} (1-e^{-At})~\text{and } \gamma_{r}\equiv \cosh\eta_{r}=\frac{1}{\sqrt{1-v_{r}^{2}}}.\nonumber\\\label{A8}	
\end{eqnarray}

The invariant momentum distribution is calculated using a local thermal distribution integrated over the freeze-out hypersurface following the Cooper--Frye prescription~\cite{Cooper:1974mv},
\begin{equation}
	E\frac{d^3N}{d^3\vec{p}}=\frac{g}{(2\pi)^3} \int_{\Sigma_{f}} f(x,p)~ p^\mu  d\Sigma_\mu,
	\label{eq:cooper_frye1}
\end{equation}
where $f$, $g$, and $\Sigma_{f}$ are the single particle distribution function, spin degeneracy factor, and the kinetic freeze-out hyper surface, respectively. The local equlibrium distrubution is given by $f(x,p) =\left[\exp\!\left(\dfrac{u^{\alpha}p_{\alpha} - \mu}{T}\right) \pm 1\right]^{-1}$, where the particle four momentum $p^\mu =\left(m_T \cosh y_{p}, p_T \cos\phi_p, p_T \sin\phi_p, m_T \sinh y_{p} \right)$ with transverse momentum  $p_{T}=\sqrt{p_{x}^{2}+p_{y}^{2}}$, transverse mass $m_{T}^{2}=p_{T}^{2}+m^{2}$, longitudinal momentum rapidity $y_{p}=\frac{1}{2}\ln \frac{E+p_{z}}{E-p_{z}}$. The momentum space azimuthal angle $\phi_{p}$ is measured in the $XY$ plane, unlike the position space angle $\phi$ (see Fig.~\eqref{fig:sphericalfireball}). The plus or minus sign in the distribution function is chosen depending on whether the particle is a fermion or a boson. To obtain the spectra, we choose an instantaneous lab time freeze-out at time $t=t_{f}$ for which $d\Sigma_\mu = (d^3x, \mathbf{0})$ and azimuthally integrate Eq.~\eqref{eq:cooper_frye1} to get,
\begin{equation}
	\frac{d^2N}{2\pi p_{T}dp_{T}dy_{p}}=\frac{g}{(2\pi)^4}\int \frac{m_{T}\cosh y_{p} ~r^{2} \sin\theta }{e^{(u^{\alpha}p_{\alpha}-\mu_{\rm kin})/T_{\rm kin}}\pm 1} dr d\theta d\phi d\phi_{p}~,\label{eq:cooper_frye2}
\end{equation}
where the inner product $u^{\alpha}p_{\alpha}=m_{T}\cosh\eta_{r}\cosh y_{p}-v_{r}\cosh\eta_{r}[\sin\theta\cos\phi(m_{T}\sinh y_{p}+p_{T}\cos\phi_{p})+p_{T}\sin\phi_{p}$ $\cos\theta]$ and $\mu_{\rm kin}$ and $T_{\rm kin}$ are respectively the kinetic freeze-out chemical potential and temperature. Eq.~\eqref{eq:cooper_frye2} serves as the master formula for the calculation of the mid-rapidity $p_{T}$ spectra and rapidity distribution displayed in the Sec.~\eqref{results}. For the evaluation of the mid-rapidity $p_{T}$ spectra, one needs to substitute $y_{p}=0$ in the RHS of Eq.~\eqref{eq:cooper_frye2}. Similarly, to see the rapidity distribution, one needs to multiply both sides of Eq.~\eqref{eq:cooper_frye2} by $2\pi p_{T}$ and integrate over the transverse momentum $dp_{T}$.

\section{RESULTS AND DISCUSSION}\label{results}
Having introduced the expanding spherical fireball model in Sec.~\eqref{modeldesc}, we now confront it with the measured transverse-momentum spectra at mid-rapidity reported by the STAR Collaboration~\cite{STAR:2017sal}. The spectra are calculated using Eq.~\eqref{eq:cooper_frye2}, with the fluid four-velocity specified by Eqs.~\eqref{A6}--\eqref{A8}.

The upper panels of Fig.~\eqref{fig:pion_pT} show the fits to the $p_{T}$ spectra of $\pi^{+}$ for different collision centralities, arranged from central (left) to peripheral collisions (right). The fitting parameters are the asymptotic radial velocity $v_{\infty}$, the radial-flow development parameter $A$, the kinetic freeze-out temperature $T_{\rm kin}$, and the freeze-out time $t_{f}$. Once these parameters are determined from the $\pi^{+}$ spectra, they are used unchanged for the $\pi^{-}$ spectra. Both pion spectra are well reproduced with vanishing chemical potential. Note that this has to do with the difference in the number of $\pi^{+}$ and $\pi^{-}$ measured in different $p_{T}$ bins. A measurable difference would result in a finite chemical potential for one or both the species. The extracted kinetic freeze-out parameters for all centralities are summarized in Table~\eqref{tab:freezeout}.

Table~\eqref{tab:freezeout} contains two independent sets of kinetic freeze-out parameters. We first discuss the results obtained from the full $p_{T}$ fitting range, denoted by the subscript ``1,'' while the second set, obtained from restricted fitting intervals, is discussed later to assess the sensitivity of the extracted parameters. For a fixed collision energy, the freeze-out time $(t_{f})_{1}$ increases from peripheral to central collisions, reflecting the longer lifetime of the larger fireball created in central events. Correspondingly, the kinetic freeze-out temperature $(T_{\rm kin})_{1}$ decreases towards central collisions, indicating that the system undergoes more substantial cooling before decoupling.
 \begin{widetext}
	
	\begin{figure}[H]
		\centering
		\begin{subfigure}{\textwidth}
			\centering
			\includegraphics[width=\linewidth]{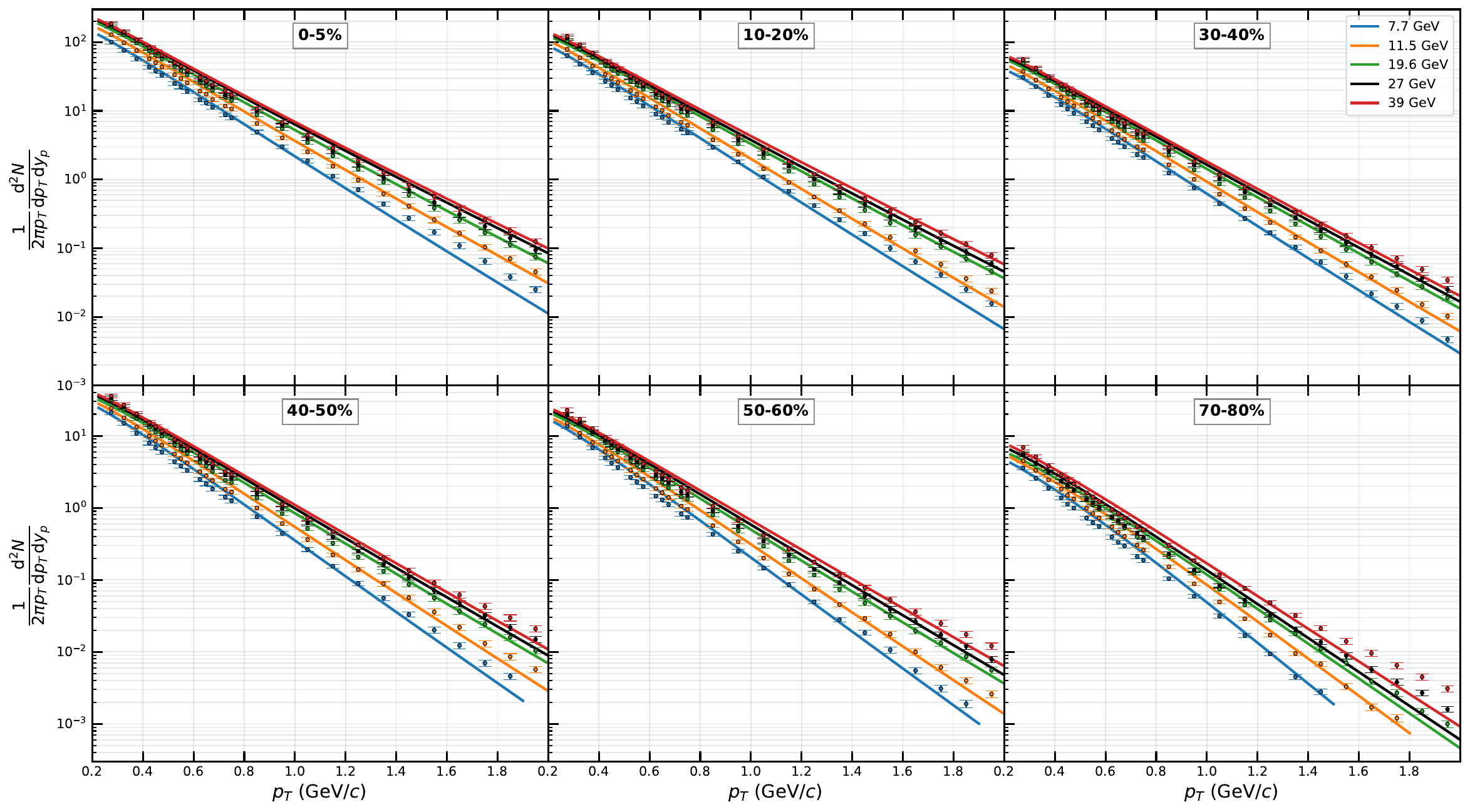}
			\label{fig:pion_plus_pT}
		\end{subfigure}
		
		\begin{subfigure}{\textwidth}
			\centering
			\includegraphics[width=\linewidth]{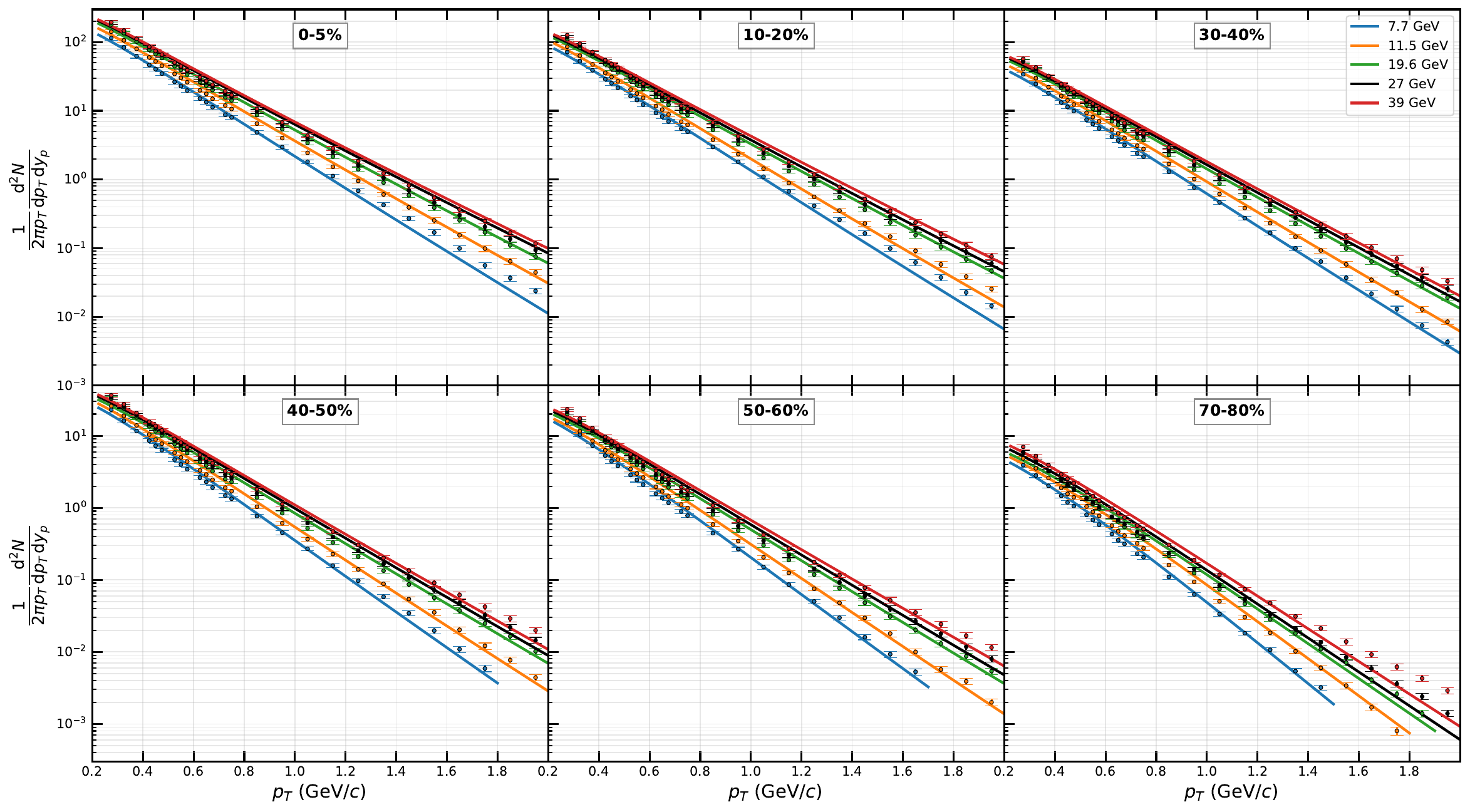}
			\label{fig:pion_minus_pT}
		\end{subfigure}
		\caption{(Color online) Calculated $p_T$ spectra of (a) $\pi^{+}$ (upper) and (b) $\pi^{-}$ (below) at $y_{p}=0$ for different centrality bins compared with STAR data~\cite{STAR:2017sal}.}
		\label{fig:pion_pT}
	\end{figure}
	\begin{figure}[H]
		\centering
		\begin{subfigure}{\textwidth}
			\centering
			\includegraphics[width=\linewidth]{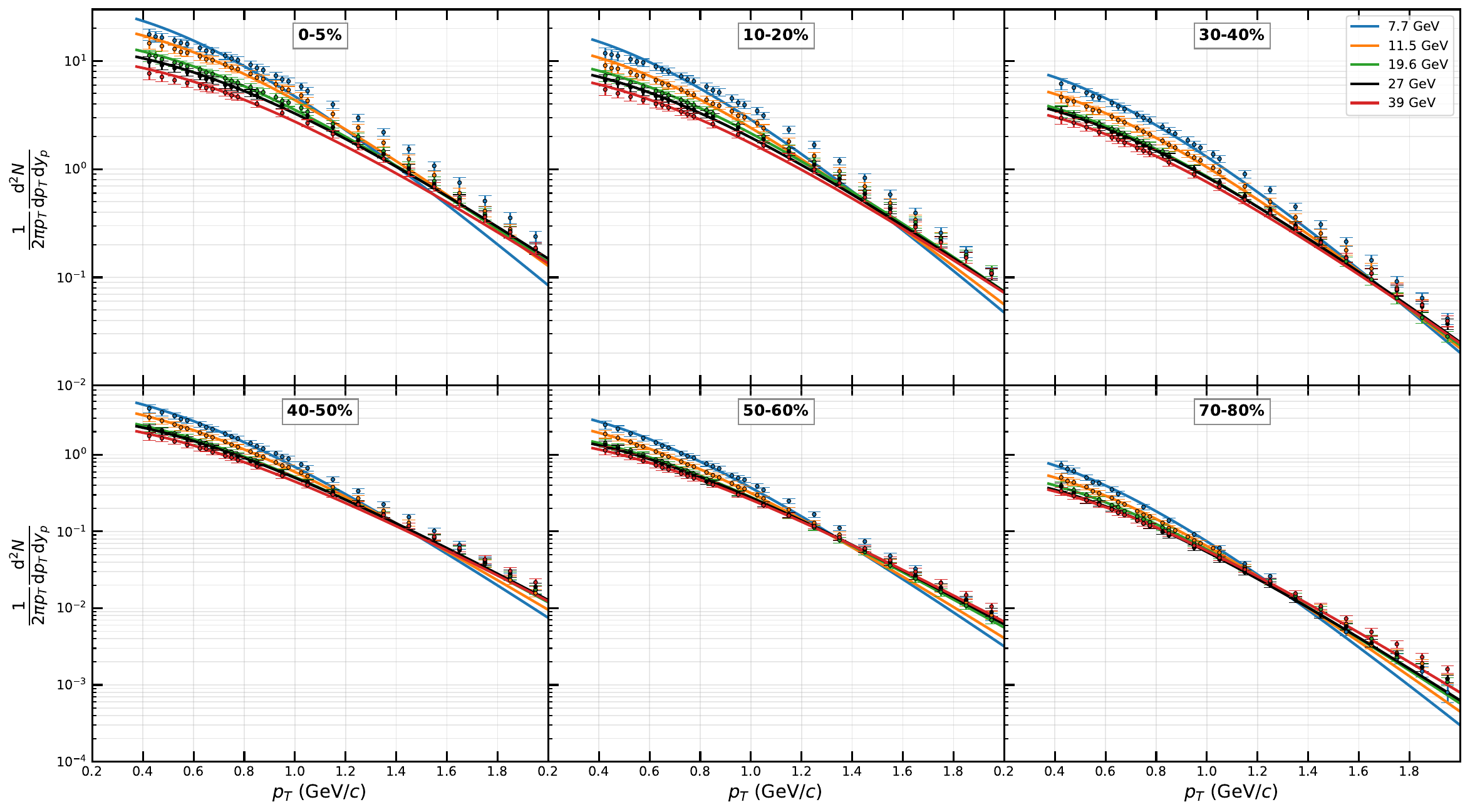}
			\label{fig:proton_pT}
		\end{subfigure}
		\begin{subfigure}{\textwidth}
			\centering
			\includegraphics[width=\linewidth]{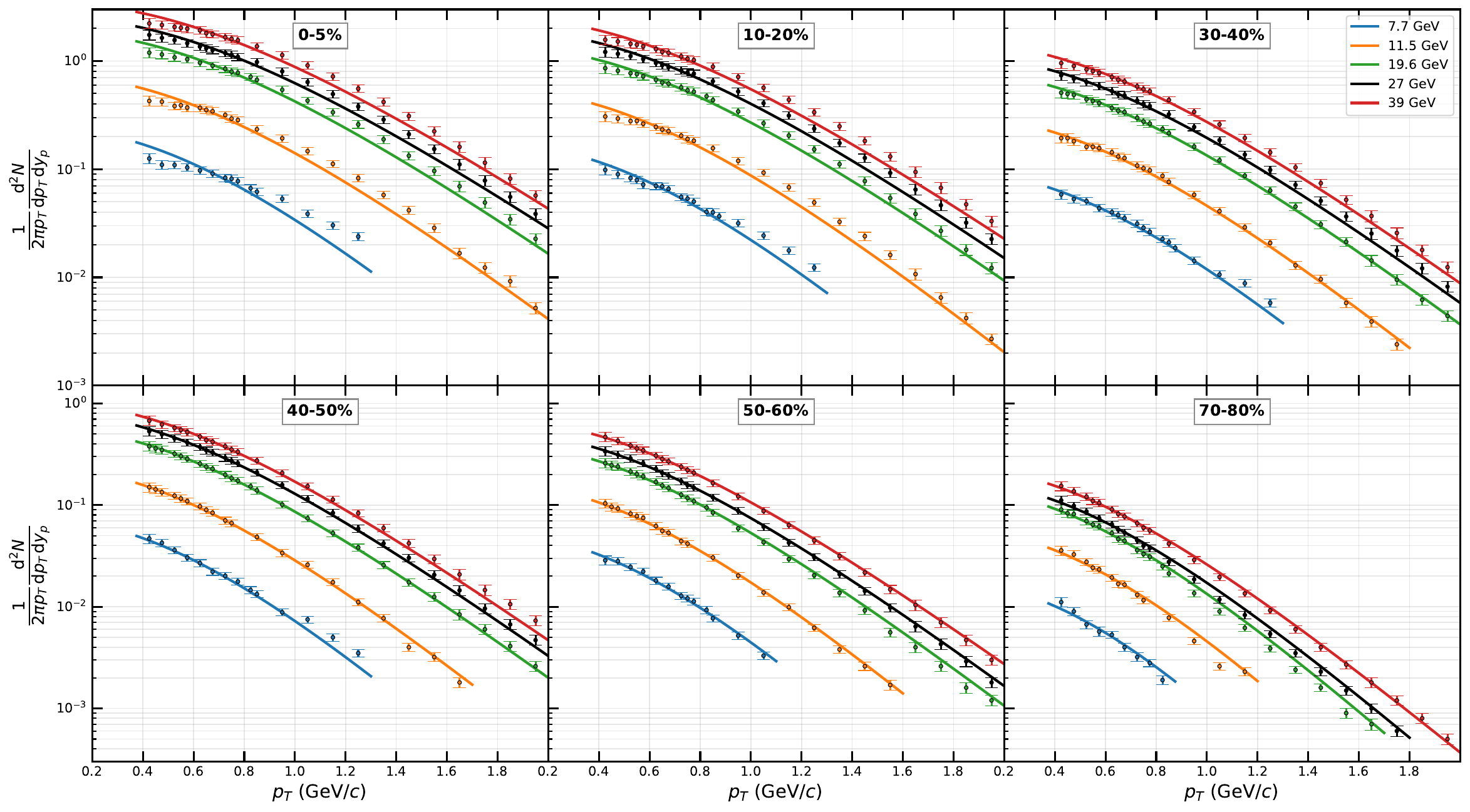}
			\label{fig:anti_proton_pT}
		\end{subfigure}
		\caption{(Color online) Calculated $p_T$ spectra of (a) $p$ (upper) and (b) $\bar{p}$ (below) at $y_{p}=0$ for different centrality bins compared with STAR data~\cite{STAR:2017sal}.}
		\label{fig:protons_pT}
	\end{figure}
	\begin{figure}[H]
		\centering
		\begin{subfigure}{\textwidth}
			\centering
			\includegraphics[width=\linewidth]{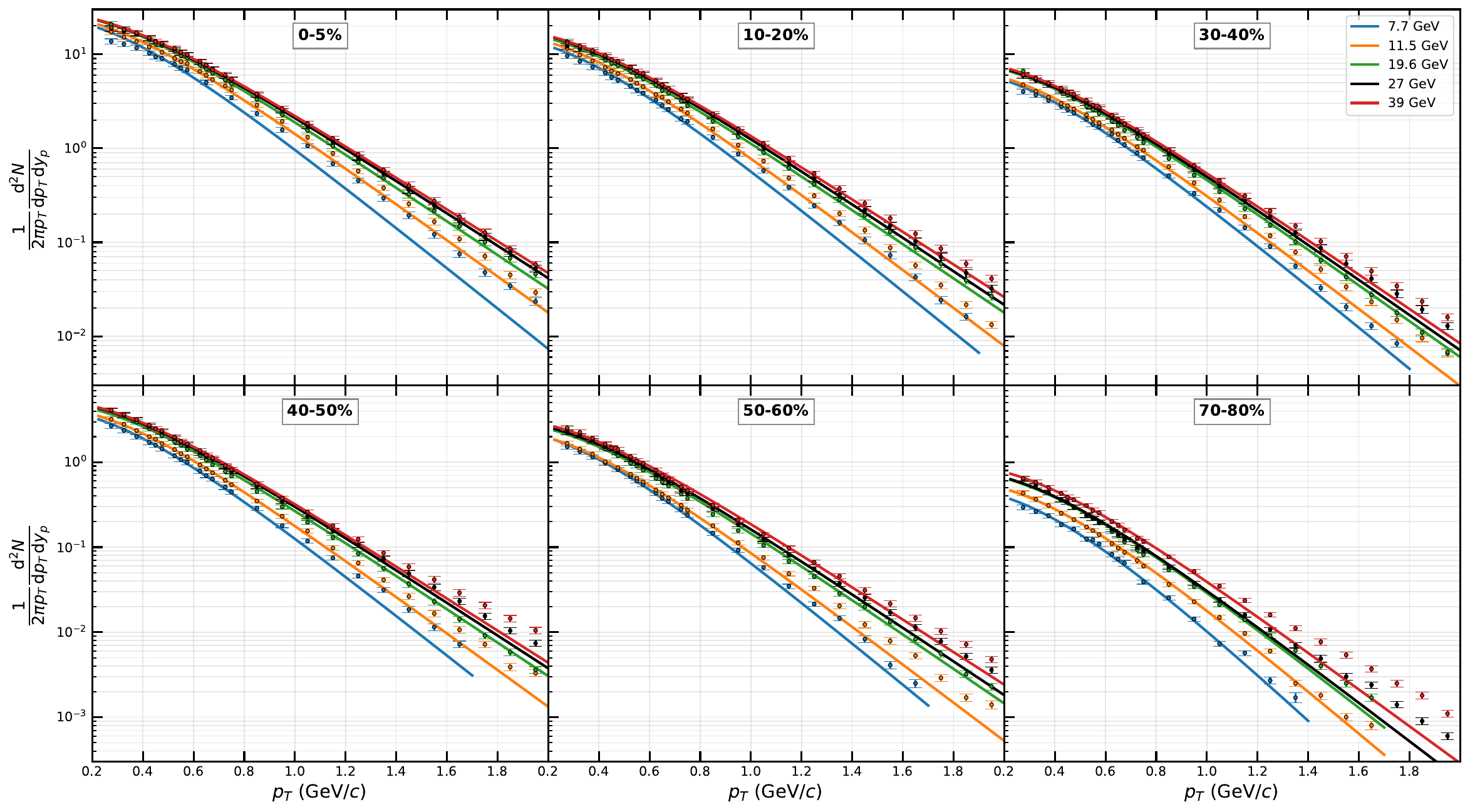}
			\label{fig:kaon_plus_pT}
		\end{subfigure}
		\begin{subfigure}{\textwidth}
			\centering
			\includegraphics[width=\linewidth]{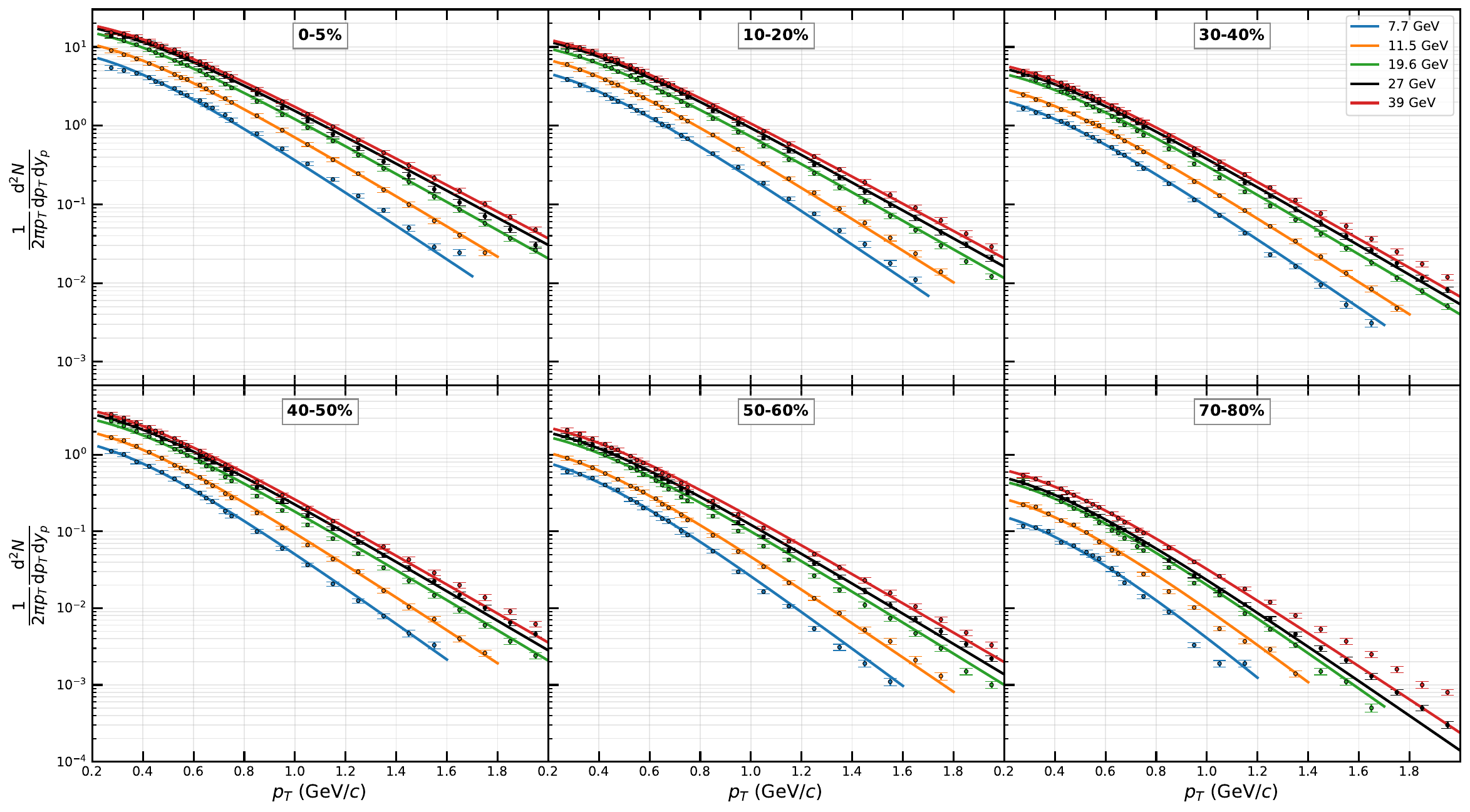}
			\label{fig:kaon_minus_pT}
		\end{subfigure}
		\caption{(Color online) Calculated transverse momentum ($p_T$) spectra of (a) $K^{+}$ (upper) and (b) $K^{-}$ (below) at mid-rapidity for different centrality bins compared with the experimental data from~\cite{STAR:2017sal}.}
		\label{fig:kaons_pT}
	\end{figure}
	\begin{figure}[H]
		\centering
		\begin{subfigure}{\textwidth}
			\centering
			\includegraphics[width=\linewidth]{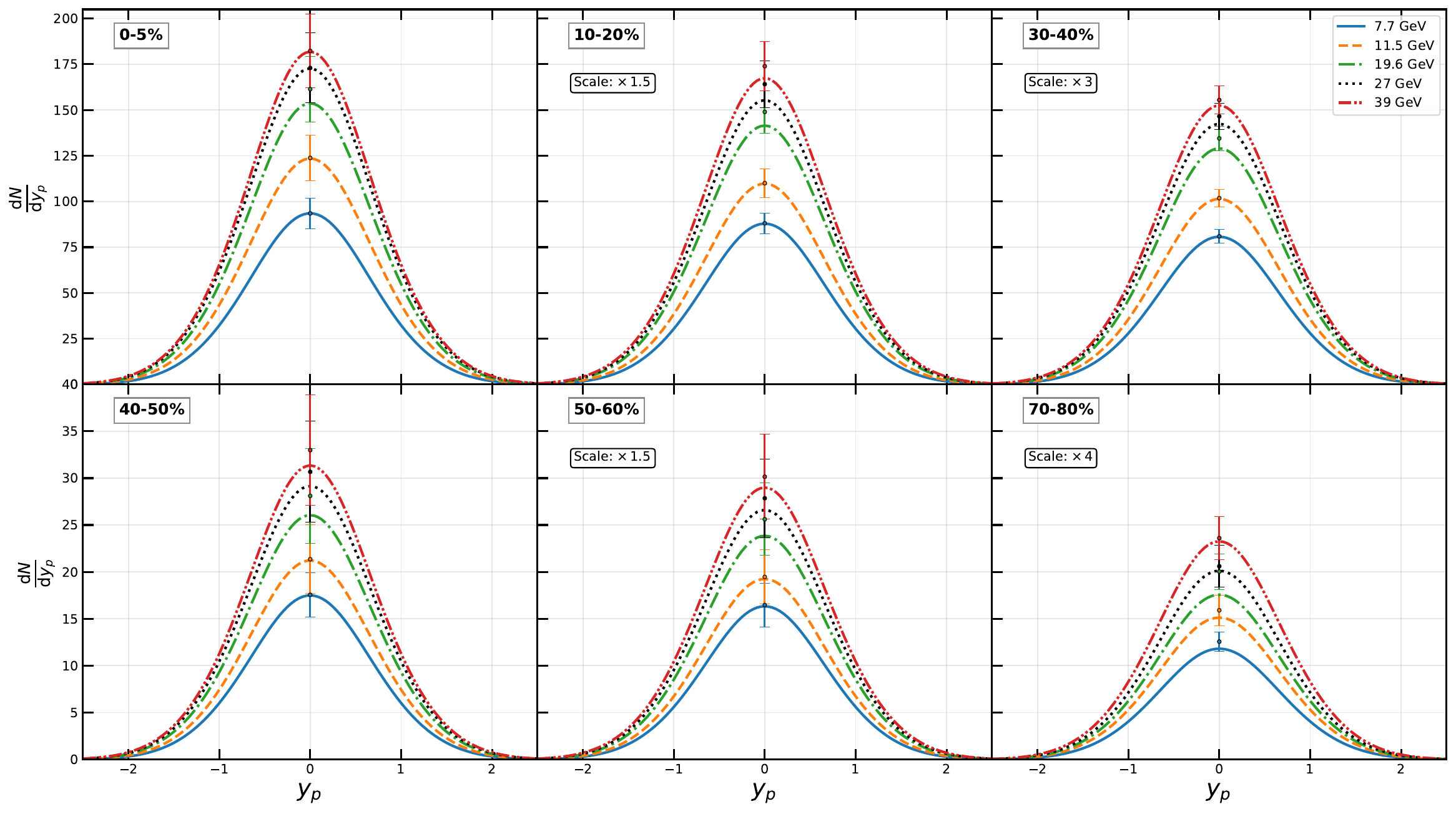}
			\label{fig:pion_plus_dNdy}	
		\end{subfigure}
		
		\begin{subfigure}{\textwidth}
			\centering
			\includegraphics[width=\linewidth]{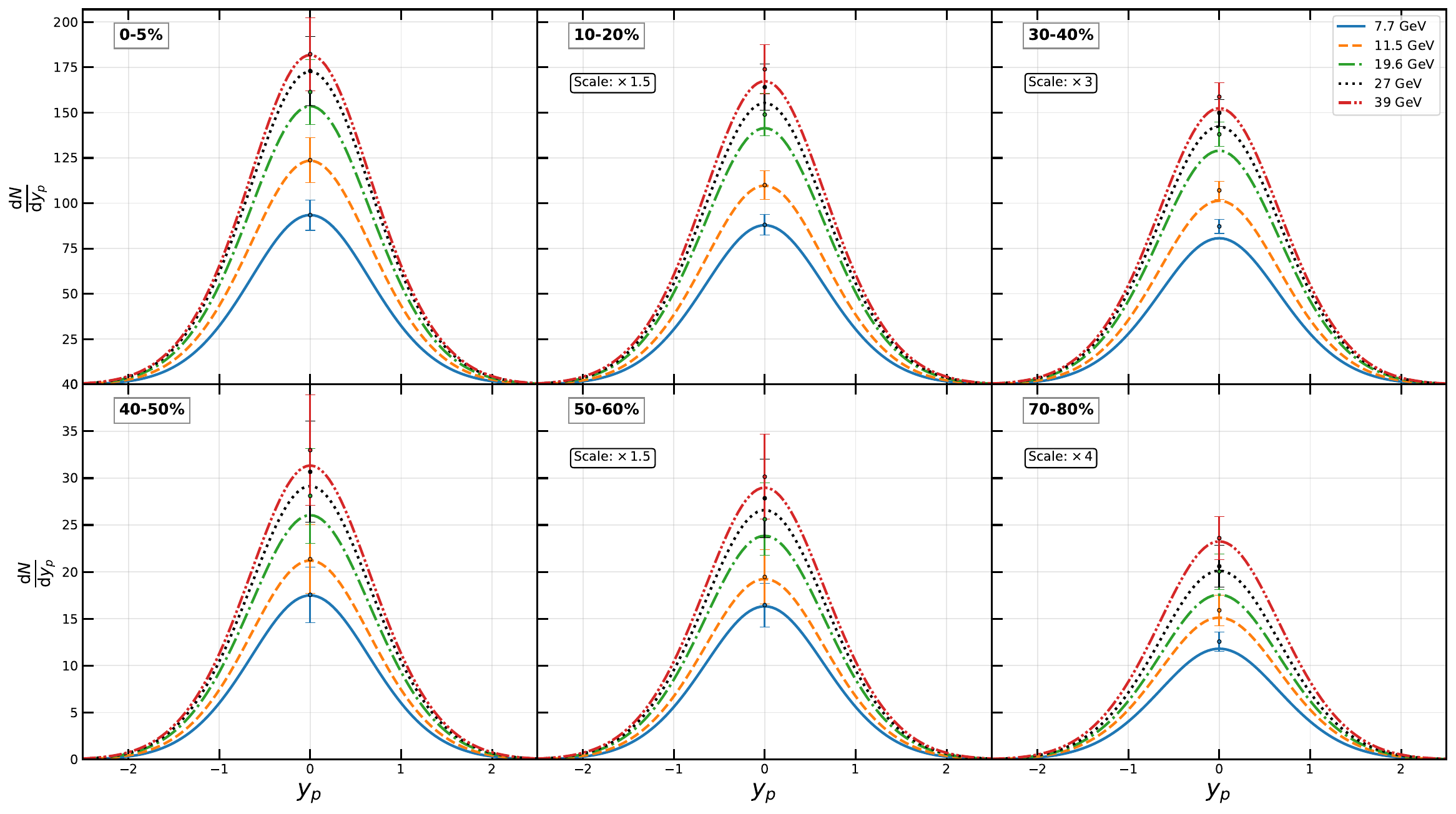}
			\label{fig:pion_minus_dNdy}	
		\end{subfigure}
		\caption{(Color online) Calculated rapidity distribution $\frac{dN}{dy_{p}}$ of (a) $\pi^{+}$ (above) and (b) $\pi^{-}$ (below) for different centrality bins compared with STAR data~\cite{STAR:2017sal} at $y_{p}=0$.}
		\label{fig:pions_dNdy}
	\end{figure}
	\begin{figure}[H]
		\centering
		\begin{subfigure}{\textwidth}
			\centering
			\includegraphics[width=\linewidth]{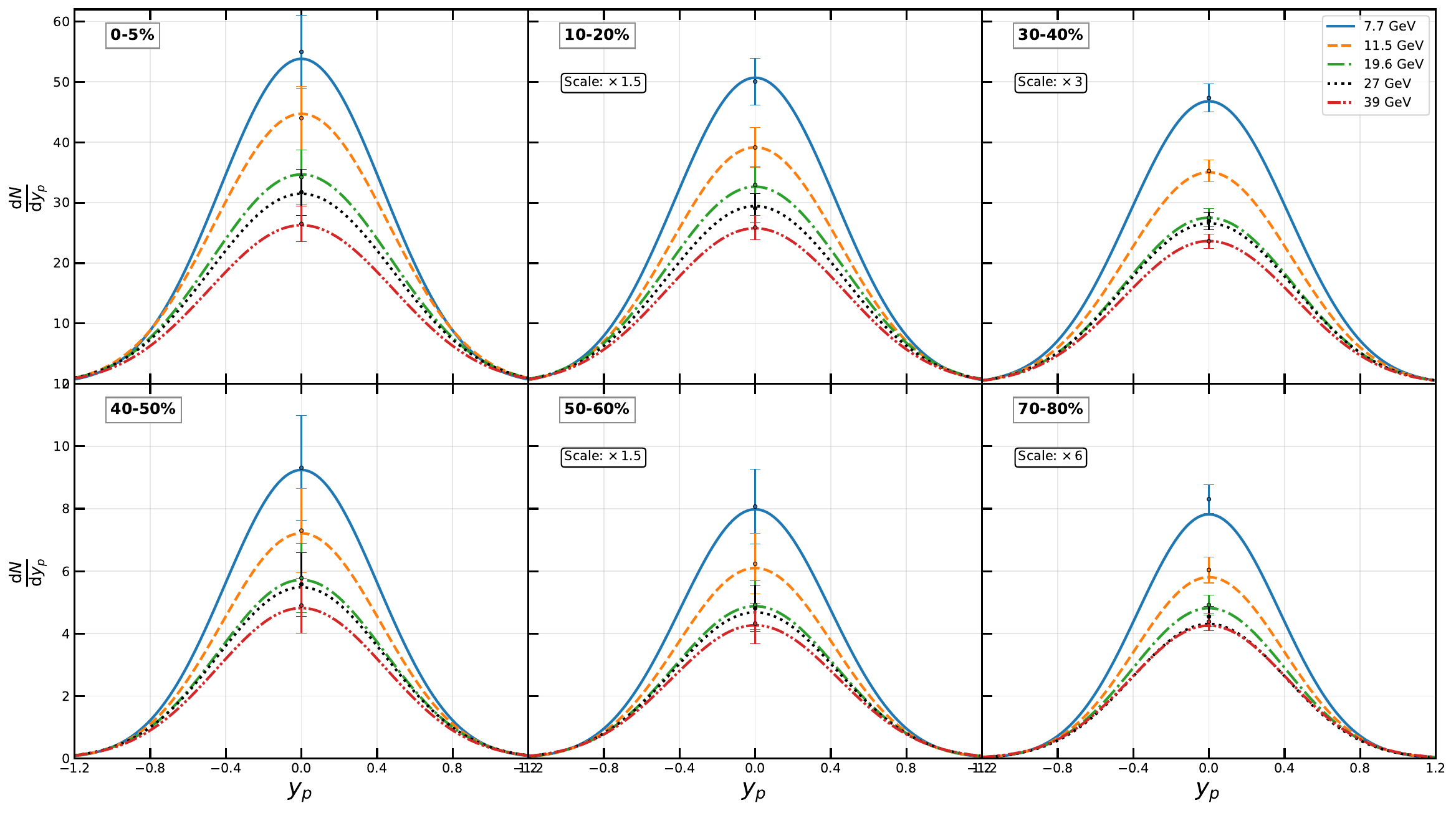}
			\label{fig:proton_dNdy}
		\end{subfigure}
		\begin{subfigure}{\textwidth}
			\centering
			\includegraphics[width=\linewidth]{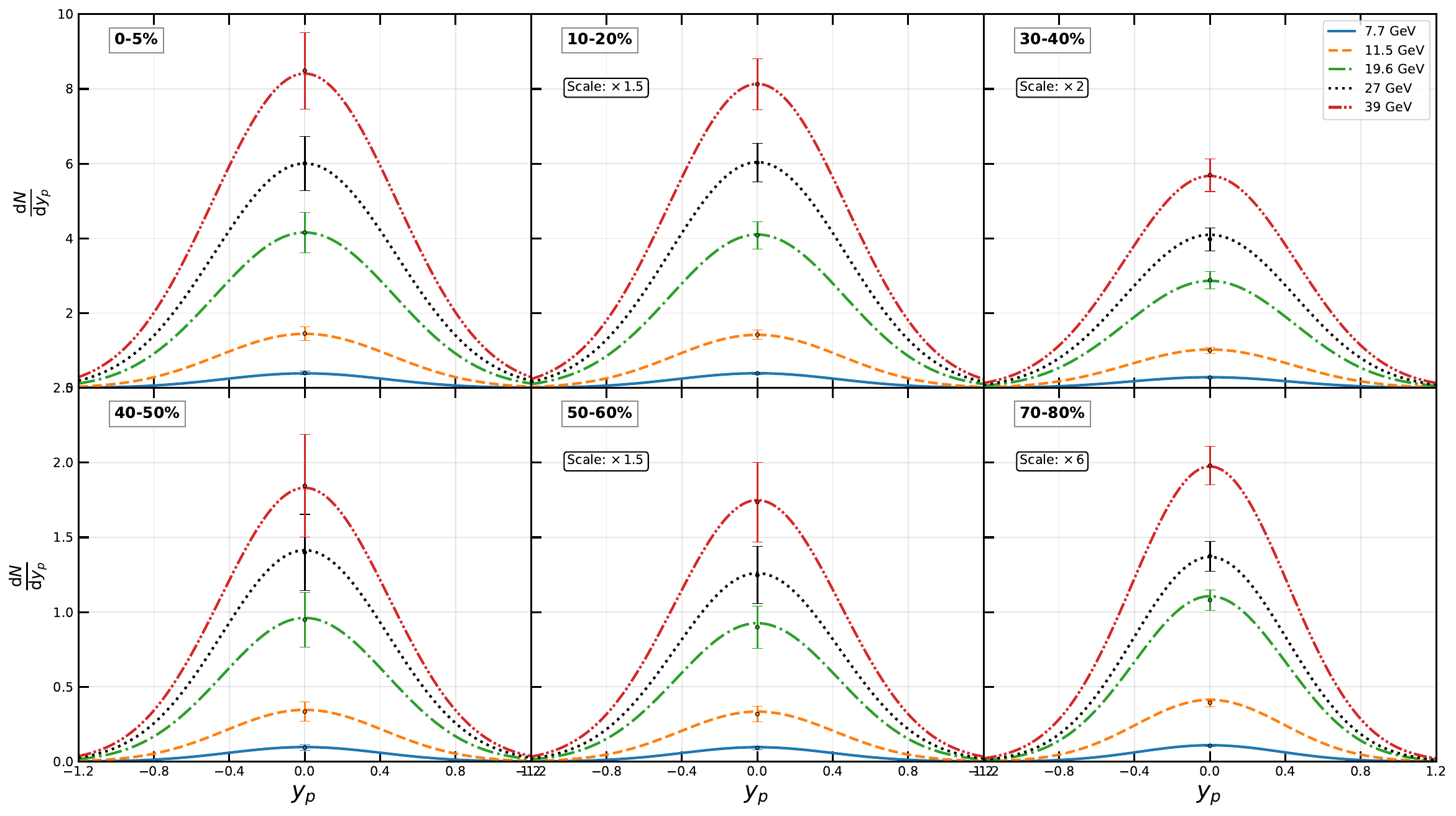}
			\label{fig:anti_proton_dNdy}
		\end{subfigure}
		\caption{(Color online) Calculated rapidity distribution $\frac{dN}{dy_{p}}$ of (a) $p$ (above) and (b) $\bar{p}$ (below) for different centrality bins compared with STAR data~\cite{STAR:2017sal} at $y_{p}=0$.}
		\label{fig:protons_dNdy}
	\end{figure}
	\begin{figure}[H]
		\centering
		\begin{subfigure}{\textwidth}
			\includegraphics[width=\linewidth]{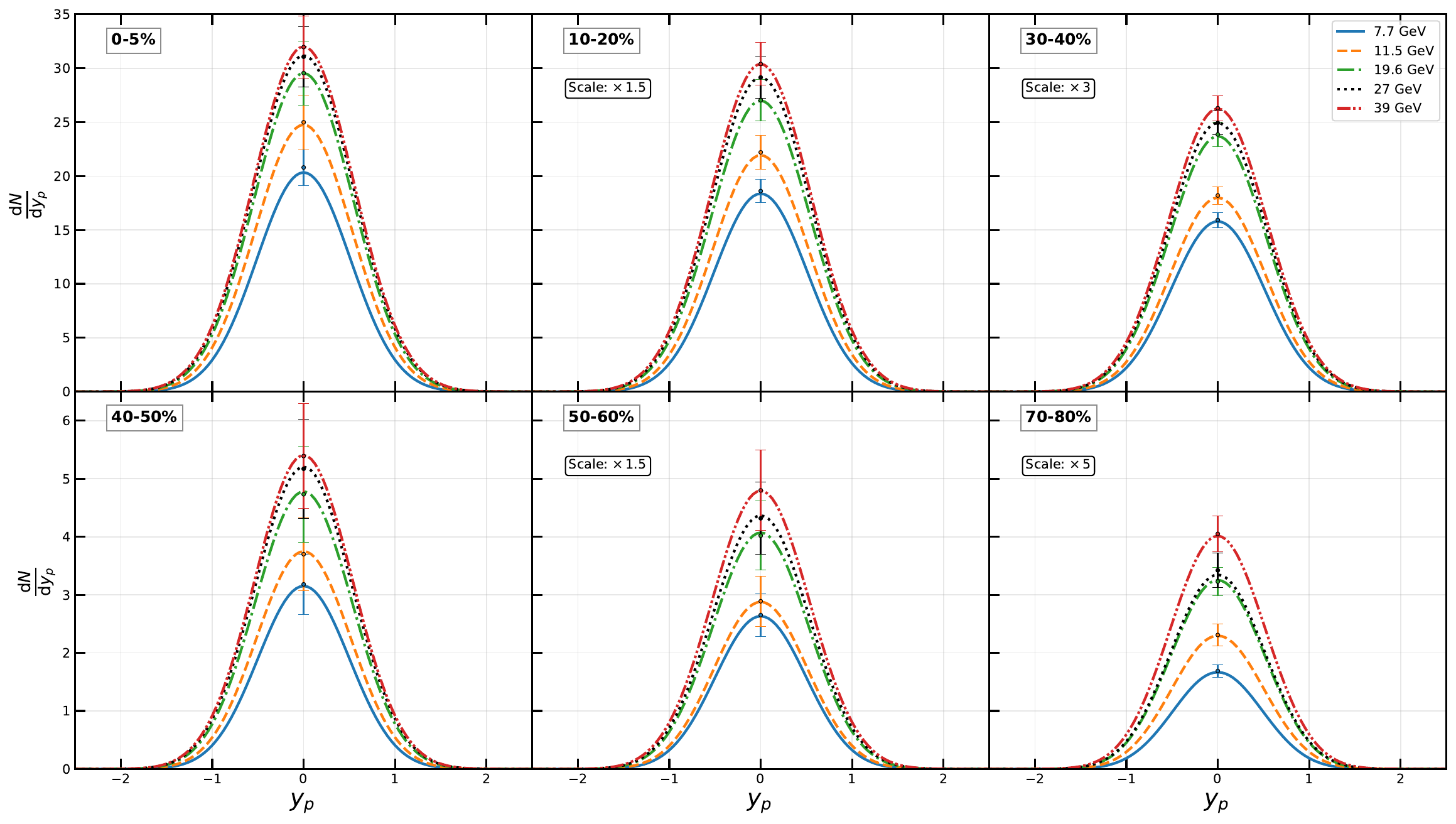}
			\label{fig:kaon_plus_dNdy}
		\end{subfigure}
		\begin{subfigure}{\textwidth}
			\centering
			\includegraphics[width=\linewidth]{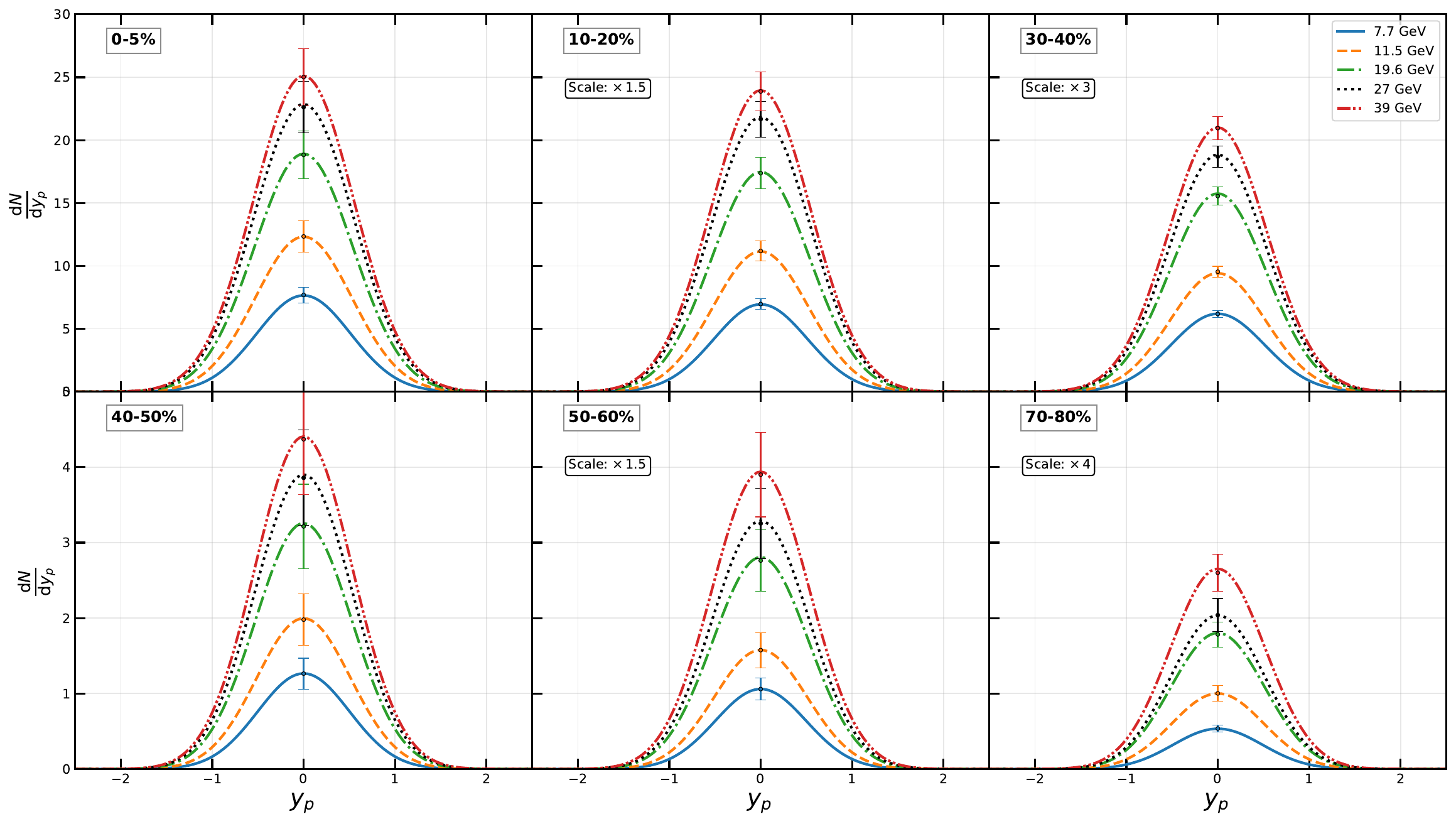}
			\label{fig:kaon_minus_dNdy}
		\end{subfigure}
		\caption{(Color online) Calculated rapidity distribution $\frac{dN}{dy_{p}}$ of (a) $K^{+}$ (above) and (b) $K^{-}$ (below) for different centrality bins compared with the data from~\cite{STAR:2017sal} at $y_{p}=0$.}
		\label{fig:kaons_dNdy}
	\end{figure}
	\begin{table}[H]
		\caption{Extracted kinetic freeze-out parameters obtained from fits of the $p_T$ spectra of $\pi^{+}$ in Au+Au collisions at different $\sqrt{s_{\rm NN}}$ and collision centralities. Parameter Set-1 and Parameter Set-2 correspond to the two fitting schemes discussed in the text. The last three columns list the corresponding STAR Blast-Wave fit results. For STAR results, the quoted errors in parenthesis are the combined statistical and systematic uncertainties. The $\sqrt{s_{\rm NN}}$, $T_{\rm kin}$, $A$, and $t_f$ are measured in units of GeV, MeV, fm$^{-1}$, and fm, respectively.}
		\label{tab:freezeout}
		\begin{ruledtabular}
			\scriptsize
			\begin{tabular}{cc|cccccc|cccccc|ccc}
				\hline
				&
				&
				\multicolumn{6}{c|}{\textbf{Parameter Set-1}}
				&
				\multicolumn{6}{c|}{\textbf{Parameter Set-2}}
				&
				\multicolumn{3}{c}{\textbf{STAR Blast-Wave~\cite{STAR:2017sal}}}\\
				
				\cline{3-8}
				\cline{9-14}
				\cline{15-17}
				
				Centrality (\%) &
				$\sqrt{s_{\rm NN}}$
				&
				$T_{\rm kin}$
				&
				$v_{\infty}$
				&
				$A$
				&
				$t_f$
				&
				$\langle v\rangle$
				&
				$\chi^2/{\rm NDF}$
				&
				$T_{\rm kin}$
				&
				$v_{\infty}$
				&
				$A$
				&
				$t_f$
				&
				$\langle v\rangle$
				&
				$\chi^2/{\rm NDF}$
				&
				$T_{\rm kin}$
				&
				$\langle\beta\rangle$
				&
				$\chi^2/{\rm NDF}$\\
				
				\hline
				
				
				\multirow{5}{*}{0--5}
				&7.7  &117.0&0.524&0.55&16.5&0.393&7.8&   127.1&0.562&0.22&14.6&0.405 &1.82 &  116(11)&0.462(0.043)&0.52\\
				&11.5 &120.0&0.586&0.65&17.0&0.440 &2.8&   123.0&0.580&0.54&15.5&0.435 &1.07 &  118(12)&0.464(0.044)&0.26\\
				&19.6 &123.0&0.620&0.75&18.0&0.465 &2.2&   126.0&0.596&0.70&17.0&0.447 &0.78 &  113(11)&0.458(0.034)&0.19\\
				&27   &124.0&0.640&0.85&18.5&0.480 &4.0&   126.9&0.609&0.75&17.5&0.457 &0.97 &  117(11)&0.482(0.038)&0.33\\
				&39   &124.5&0.650&0.90&19.0&0.487 &3.1&   126.9&0.627&0.80&18.0&0.470 &0.91 &  117(11)&0.492(0.038)&0.18\\
				
				\hline
				
				
				\multirow{5}{*}{10--20}
				&7.7  &119.0&0.502&0.55&13.5&0.376 &5.7&   124.7&0.550&0.43&10.7&0.408 &0.73 &  121(12)&0.403(0.040)&0.39\\
				&11.5 &121.0&0.547&0.60&14.5&0.410 &3.2&   123.4&0.570&0.51&12.6&0.427 &0.57 &  120(12)&0.423(0.038)&0.23\\
				&19.6 &129.0&0.580&0.65&15.0&0.435 &3.0&   126.6&0.580&0.60&14.5&0.435 &0.51 &  117(12)&0.435(0.032)&0.30\\
				&27   &130.0&0.590&0.80&15.5&0.442 &2.8&   129.8&0.590&0.65&14.7&0.442 &0.66 &  120(11)&0.452(0.028)&0.46\\
				&39   &132.0&0.600&0.80&16.0&0.450 &3.1&   131.9&0.600&0.70&15.0&0.450 &0.59 &  120(11)&0.456(0.034)&0.15\\
				
				\hline
				
				
				\multirow{5}{*}{30--40}
				&7.7  &120.6&0.490&0.45&10.0&0.363 &2.9&  126.2&0.510&0.479&7.6&0.372 &1.05 &     129(12)&0.348(0.049)&0.61\\
				&11.5 &124.3&0.520&0.50&11.0&0.388 &3.0&   127.4&0.520&0.500&9.4&0.386 &0.35 &    133(13)&0.363(0.056)&0.22\\
				&19.6 &132.0&0.530&0.55&12.0&0.397 &3.5&   128.9&0.530&0.550&11.4&0.397 &0.46 &   123(12)&0.360(0.026)&0.40\\
				&27   &133.0&0.540&0.65&12.5&0.405 &3.3&   132.2&0.540&0.600&11.7&0.405 &0.33 &   131(10)&0.381(0.029)&0.28\\
				&39   &134.0&0.550&0.70&13.0&0.412 &3.6&   134.4&0.550&0.650&12.0&0.412 &0.34 &   129(11)&0.394(0.033)&0.11\\
				
				\hline
				
				
				\multirow{5}{*}{40--50}
				&7.7  &121.0&0.435&0.40&9.5&0.319 &4.3&    129.7&0.450&0.45&6.5&0.319 &1.29 &   131(12)&0.282(0.044)&0.74\\
				&11.5 &127.0&0.464&0.45&10.0&0.318 &3.2&   131.7&0.470&0.51&7.9&0.346 &0.56 &   136(13)&0.271(0.034)&0.27\\
				&19.6 &135.0&0.490&0.50&10.5&0.343 &3.5&   133.3&0.480&0.52&10.0&0.358 &0.58 &  129(13)&0.315(0.024)&0.39\\
				&27   &136.0&0.500&0.60&11.0&0.352 &3.4&   135.6&0.490&0.55&10.5&0.366 &0.40 &  133(10)&0.324(0.027)&0.22\\
				&39   &137.0&0.510&0.65&11.5&0.367 &3.7&   138.8&0.500&0.60&10.7&0.374 &0.45 &  131(12)&0.345(0.031)&0.11\\
				
				\hline
				
				
				\multirow{5}{*}{50--60}
				&7.7  &122.0&0.409&0.30&8.9&0.286 &4.3&   131.6&0.440&0.28&6.0&0.270 &1.76 &  139(13)&0.205(0.053)&1.25\\
				&11.5 &129.0&0.428&0.35&9.2&0.308 &3.9&   132.0&0.450&0.40&7.1&0.317 &0.62&   139(14)&0.207(0.033)&0.33\\
				&19.6 &137.0&0.460&0.40&9.6&0.337 &4.8&   134.1&0.460&0.42&8.9&0.337 &0.77 &  132(13)&0.246(0.026)&0.31\\
				&27   &138.0&0.470&0.50&9.9&0.350 &4.6&   135.5&0.470&0.47&9.4&0.348 &0.41 &  139(10)&0.253(0.028)&0.13\\
				&39   &139.0&0.490&0.55&10.2&0.366 &4.1&  139.4&0.480&0.48&9.7&0.356 &0.52 &  138(13)&0.277(0.028)&0.10\\
				
				\hline
				
				
				\multirow{5}{*}{70--80}
				&7.7  &128.0&0.250&0.20&8.1&0.150 &5.1&   134.0&0.243&0.25&5.0&0.130 &3.08 &  140(13)&0.106(0.035)&0.89\\
				&11.5 &136.5&0.280&0.25&8.5&0.185 &6.2&   130.9&0.390&0.42&5.3&0.261 &2.05 &  140(14)&0.147(0.032)&0.78\\
				&19.6 &142.0&0.300&0.30&9.1&0.210 &4.6&   135.0&0.400&0.43&6.6&0.283 &1.63 &  137(14)&0.174(0.028)&1.11\\
				&27   &139.0&0.348&0.35&9.2&0.251 &7.2&   137.5&0.410&0.44&6.9&0.293 &1.29 &  142(11)&0.176(0.029)&1.01\\
				&39   &140.0&0.380&0.37&9.5&0.276 &7.3&   143.9&0.420&0.45&7.2&0.303 &1.70 &  143(12)&0.208(0.022)&0.39\\
				
				\hline
			\end{tabular}
		\end{ruledtabular}
	\end{table}
\end{widetext}
 Furthermore, the fitted values of $(A)_{1}$ show that the radial flow develops considerably faster in central than in peripheral collisions, consistent with the stronger pressure gradients expected in larger collision systems.

The flow parameters $(A)_{1}$, $(v_{\infty})_{1}$, and $(t_{f})_{1}$ extracted from the pion spectra are subsequently used without modification to calculate the $p_{T}$ spectra of protons and kaons. The only remaining free parameter is the particle-dependent kinetic freeze-out chemical potential, which is adjusted to reproduce the measurements of STAR collaboration~\cite{STAR:2017sal}. The resulting spectra are compared with the experimental data in Figs.~\eqref{fig:protons_pT} and \eqref{fig:kaons_pT}, while the extracted values of $(\mu_{\rm kin})_{1}$ are listed in Table~\eqref{tab:chemical_potentials}.

The corresponding rapidity distributions of identified hadrons are presented in Figs.~\eqref{fig:pions_dNdy}--\eqref{fig:kaons_dNdy}. These distributions are calculated using the first set of kinetic freeze-out parameters given in Table~\eqref{tab:freezeout}. The model predicts approximately Gaussian rapidity distributions, consistent with observations at low and intermediate collision energies reported by the CERN SPS and GSI experiments~\cite{NA49:2002pzu,HADES:2020ver}. Such Gaussian-like behavior has also been observed up to $\sqrt{s_{\rm NN}}=200$ GeV~\cite{Ouerdane:2002gm,PhysRevLett.93.102301,PhysRevLett.94.162301}. Since STAR collaboration measurements are available only at mid-rapidity, only this point is shown in Figs.~\eqref{fig:pions_dNdy}--\eqref{fig:kaons_dNdy}.

Our approach differs from the conventional statistical thermal analysis of particle yields, where either the mid-rapidity yields, $\frac{dN}{dy_{p}}\vert_{y_{p}=0}$, or the total $4\pi$ yields, $\int \frac{dN}{dy_{p}}dy_{p}$,
are analyzed within statistical thermal models to extract the chemical freeze-out temperature and chemical potentials~\cite{Cleymans:1998fq,Cleymans:1999st,Becattini:2000jw,Braun-Munzinger:2003pwq,Cleymans:2005xv,Wheaton_2009,Andronic:2005yp,Andronic:2017pug,Chatterjee:2015fua,Chatterjee:2017yhp}. The success of these analyses relies on two important observations. First, particle-yield ratios obtained from the integrated $4\pi$ yields are insensitive to the collective expansion of the fireball and depend primarily on the thermodynamic conditions at chemical freeze-out~\cite{Cleymans:1999st,FLORKOWSKI2025100249}. Second, under the assumption of longitudinal boost invariance~\cite{Bjorken:1982qr}, the rapidity distributions 
\begin{table*}[t]
	\caption{Extracted chemical potentials obtained from the fits of the $p_{T}$ spectra of $p$, $\bar{p}$, $K^{+}$ and $K^{-}$ in Au+Au collisions at different $\sqrt{s_{\rm NN}}$ and collision centralities. All chemical potentials are given in MeV.}
	\label{tab:chemical_potentials}
	\begin{ruledtabular}
		\scriptsize
		\begin{tabular}{cc|cccc|cccc}
			\hline
			&
			&
			\multicolumn{4}{c|}{\textbf{Parameter Set-1}}
			&
			\multicolumn{4}{c}{\textbf{Parameter Set-2}}\\
			
			\cline{3-6}
			\cline{7-10}
			
			Centrality (\%) &
			$\sqrt{s_{\rm NN}}$
			&
			$\mu_{p}$
			&
			$\mu_{\bar p}$
			&
			$\mu_{K^{+}}$
			&
			$\mu_{K^{-}}$
			&
			$\mu_{p}$
			&
			$\mu_{\bar p}$
			&
			$\mu_{K^{+}}$
			&
			$\mu_{K^{-}}$\\
			
			\hline
			
			
			\multirow{5}{*}{0--5}
			&7.7  &374.6&-202.6&9.9&-104.0&   377.5&-249.8&18.7&-111.9\\
			&11.5 &315.3&-96.7&-4.1&-87.5&    331.9&-93.4&14.8&-74.9\\
			&19.6 &250.1&-10.9&-14.8&-69.4&   260.0&-9.2&-1.5&-59.2\\
			&27   &221.2&16.1&-24.0&-62.3&   233.0&22.6&-5.9&-47.7\\
			&39   &191.5&49.7&-27.9&-58.0&    202.0&56.6&-10.5&-41.8\\
			
			\hline
			
			
			\multirow{5}{*}{10--20}
			&7.7  &367.6&-211.6&-0.1&-115.4&  387.5&-220.7&16.9&-110.4\\
			&11.5 &307.2&-94.4&-8.4&-89.9&    328.9&-80.4&13.2&-74.0\\
			&19.6 &227.7&-40.0&-31.5&-87.3&   257.6&-4.1&-5.4&-61.9\\
			&27   &199.3&-6.7&-35.5&-72.9&    226.2&19.3&-11.7&-52.3\\
			&39   &165.0&12.8&-44.1&-75.3&    191.4&40.0&-18.4&-50.9\\
			
			\hline
			
			
			\multirow{5}{*}{30--40}
			&7.7  &362.5&-203.9&-12.3&-124.7&   380.2&-210.6&6.4&-122.4\\
			&11.5 &290.2&-98.4&-31.9&-112.0&    310.7&-86.2&-7.4&-95.1\\
			&19.6 &202.4&-42.8&-45.4&-99.0&     238.5&-1.3&-17.6&-71.2\\
			&27   &181.9&-13.0&-53.7&-90.7&     211.2&21.8&-25.1&-64.7\\
			&39   &154.0&16.9&-57.9&-87.6&      181.4&44.1&-30.6&-61.8\\
			
			\hline
			
			
			\multirow{5}{*}{40--50}
			&7.7  &346.1&-206.1&-25.1&-135.1&     360.6&-229.4&-9.0&-142.1\\
			&11.5 &463.7&-111.0&-41.0&-120.7&     294.0&-105.3&-18.8&-108.6\\
			&19.6 &192.2&-48.8&-54.1&-105.6&      222.2&-15.3&-29.6&-84.6\\
			&27   &168.4&-16.1&-59.7&-98.7&       194.3&12.7&-34.6&-76.6\\
			&39   &137.7&5.1&-66.4&-94.2&         157.0&23.1&-45.1&-76.0\\
			
			\hline
			
			
			\multirow{5}{*}{50--60}
			&7.7  &331.0&-208.9&-42.3&-152.9&     344.5&-241.8&-27.2&-163.9\\
			&11.5 &256.4&-118.6&-68.7&-146.0&     281.2&-102.1&-39.3&-126.5\\
			&19.6 &172.6&-55.3&-69.8&-120.5&      208.8&-13.7&-40.9&-95.5\\
			&27   &149.3&-31.9&-77.0&-115.9&      184.7&8.4&-44.6&-85.6\\
			&39   &122.1&-1.9&-77.7&-104.8&       149.0&24.3&-52.4&-81.9\\
			
			\hline
			
			
			\multirow{5}{*}{70--80}
			&7.7  &292.2&-255.0&-106.4&-223.4&     308.5&-265.3&-86.2&-226.4\\
			&11.5 &194.1&-167.0&-119.8&-202.4&     244.7&-107.9&-71.&-165.0\\
			&19.6 &128.6&-80.4&-106.0&-157.9&      178.6&-27.8&-69.7&-125.1\\
			&27   &108.2&-51.3&-112.2&-150.1&      147.6&-14.9&-74.4&-113.3\\
			&39   &84.0&-23.6&-108.8&-135.4&       170.7&57.4&-11.4&-40.7\\
			
			\hline
		\end{tabular}
	\end{ruledtabular}
\end{table*}
become independent of $y_{p}$, implying that the mid-rapidity yield ratios are likewise determined solely by the chemical freeze-out parameters~\cite{Cleymans:1999st,Broniowski:2001we}. 

A common ingredient of statistical thermal models is the inclusion of resonance feed-down in order to reproduce the observed inclusive particle multiplicities. Likewise, the default BW analyses of identified-hadron $p_{T}$ spectra are performed in momentum intervals where resonance decay contributions are expected to be minimal~\cite{STAR:2008med,STAR:2017sal}. As discussed in Ref.~\cite{STAR:2008med}, resonance feed-down affects the normalization of the spectra much more strongly than their shape, except for pions where the effect is more pronounced. Within the Maxwell--Boltzmann approximation, the overall normalization is governed by the fugacity factor $e^{\mu_{\rm kin}/T_{\rm kin}}$. Consequently, the extracted particle-dependent values of $\mu_{\rm kin}$ should be interpreted as effective kinetic chemical potentials that effectively absorb the missing feed-down contributions.

Following the philosophy of the default BW analysis, our first set of fits was performed over the interval $0.25~{\rm GeV}<p_{T}<2.0~{\rm GeV}$ (or the available experimental range whichever was larger), without explicitly including resonance feed-down, as shown in Figs.~\eqref{fig:pion_pT}--\eqref{fig:kaons_pT}. Greater weight was assigned to the low-$p_{T}$ region during the fitting procedure to ensure a reliable description of the experimentally measured mid-rapidity yields. Since the spectra decrease exponentially with increasing $p_{T}$, small discrepancies in the intermediate- or high-$p_{T}$ region have  only a limited  impact on the $p_{T}$ integrated rapidity distributions. It is obvious from the expression, $
\frac{dN}{dy_{p}}\vert_{y_{p}=0}
= \int \left(\frac{d^{2}N}{2\pi p_{T}dp_{T}dy_{p}}\right)_{y_{p}=0} 2\pi p_{T} dp_{T}$ that the calculated mid-rapidity yield, obtained by integrating the fitted $p_{T}$ spectrum, agrees well with the experimental data, only if the integrand agrees with the experimental data. 

To assess the sensitivity of the extracted freeze-out parameters to the fitting interval, we repeated the analysis using the momentum ranges adopted in the STAR BW study~\cite{STAR:2017sal}, where resonance and hard-process contributions are expected to be less significant. For pions, the fit was restricted to $0.5~{\rm GeV}<p_{T}<1.3~{\rm GeV}$, yielding the second set of kinetic freeze-out parameters, $(A)_{2}$, $(v_{\infty})_{2}$, and $(t_{f})_{2}$, listed in Table~\eqref{tab:freezeout}. The pion chemical potentials were again fixed to zero, except for the $\pi^{-}$ at $\sqrt{s_{\rm NN}}=39$ GeV and centrality class $70-80$\% where it is taken to be $82$ MeV. 
Comparing the parameter set-1 with 2, we notice that there is an overall mild decrease in $t_{f}$ and $A$ in set-2, which imply a lesser fireball lifetime. The same procedure was subsequently applied to kaons and protons, using the momentum intervals $0.25<p_{T}<1.4$ GeV and $0.4<p_{T}<1.3$ GeV, respectively, while varying only the effective kinetic freeze-out chemical potentials. The resulting values of $(\mu_{\rm kin})_{2}$ are given in Table~\eqref{tab:chemical_potentials}. As expected, restricting the fits to these momentum intervals leads to a noticeable improvement in the quality of the fits, as reflected by the corresponding $\chi^{2}$ values.

\section{SUMMARY}\label{summary}
In this work, we have investigated the transverse-momentum spectra and rapidity distributions of identified light hadrons produced in Au+Au collisions at RHIC Beam Energy Scan energies by using an expanding spherical fireball model. The particle spectra were calculated using the Cooper--Frye freeze-out prescription with a blast-wave-like radial flow profile, where the expansion velocity was determined self-consistently from the time evolution of the fireball radius. The kinetic freeze-out parameters were extracted by fitting the pion $p_{T}$ spectra for different collision centralities, and the same collective flow parameters were subsequently employed to describe the spectra of kaons and protons, introducing only particle-dependent effective kinetic freeze-out chemical potentials.

The model provides a satisfactory description of the measured $p_{T}$ spectra of $\pi^{\pm}$, $K^{\pm}$, $p$, and $\bar{p}$ over the considered collision energies and centralities. The extracted freeze-out parameters exhibit systematic centrality dependence: the freeze-out time increases from peripheral to central collisions, while the kinetic freeze-out temperature decreases, indicating a longer-lived and more strongly cooled fireball in central events. The radial flow develops more rapidly in central collisions, consistent with stronger collective expansion driven by larger pressure gradients. The calculated rapidity distributions exhibit approximately Gaussian shapes, in qualitative agreement with the behavior observed in heavy-ion collisions over a broad range of beam energies.

The present study demonstrates that a simple dynamical fireball model with a small number of physically motivated parameters can simultaneously account for the transverse-momentum spectra and predict the corresponding rapidity distributions in a consistent framework.

\section{ACKNOWLEDGEMENT}
This work was supported in part by the Board of Research in Nuclear Sciences (BRNS) and the Department of Atomic Energy (DAE), Government of India, with Grant Nos. 57/14/01/2024-BRNS/313 (S.G.) and the Ministry of Education, Government of India (A.D.). The authors thank Anand Rai for fruitful discussion. 
\bibliographystyle{unsrturl}
\bibliography{ref}

@article{Roland:2014jsa,
	author = "Roland, G. and Safarik, K. and Steinberg, P.",
	title = "{Heavy-ion collisions at the LHC}",
	doi = "10.1016/j.ppnp.2014.05.001",
	journal = "Prog. Part. Nucl. Phys.",
	volume = "77",
	pages = "70--127",
	year = "2014"
}

@article{Foka:2016vta,
	author = "Foka, Panagiota and Janik, Ma{\l}gorzata Anna",
	title = "{An overview of experimental results from ultra-relativistic heavy-ion collisions at the CERN LHC: Bulk properties and dynamical evolution}",
	eprint = "1702.07233",
	archivePrefix = "arXiv",
	primaryClass = "hep-ex",
	doi = "10.1016/j.revip.2016.11.002",
	journal = "Rev. Phys.",
	volume = "1",
	pages = "154--171",
	year = "2016"
}

@article{Andronic:2017pug,
    author = "Andronic, Anton and Braun-Munzinger, Peter and Redlich, Krzysztof and Stachel, Johanna",
    title = "{Decoding the phase structure of QCD via particle production at high energy}",
    eprint = "1710.09425",
    archivePrefix = "arXiv",
    primaryClass = "nucl-th",
    doi = "10.1038/s41586-018-0491-6",
    journal = "Nature",
    volume = "561",
    number = "7723",
    pages = "321--330",
    year = "2018"
}

@article{Shuryak:2003xe,
	author = "Shuryak, Edward",
	editor = "Faessler, A.",
	title = "{Why does the quark gluon plasma at RHIC behave as a nearly ideal fluid?}",
	eprint = "hep-ph/0312227",
	archivePrefix = "arXiv",
	doi = "10.1016/j.ppnp.2004.02.025",
	journal = "Prog. Part. Nucl. Phys.",
	volume = "53",
	pages = "273--303",
	year = "2004"
}

@article{Gale:2013da,
	author = "Gale, Charles and Jeon, Sangyong and Schenke, Bjoern",
	title = "{Hydrodynamic Modeling of Heavy-Ion Collisions}",
	eprint = "1301.5893",
	archivePrefix = "arXiv",
	primaryClass = "nucl-th",
	doi = "10.1142/S0217751X13400113",
	journal = "Int. J. Mod. Phys. A",
	volume = "28",
	pages = "1340011",
	year = "2013"
}

@article{De:2022yxq,
	author = "De, A. and Kapusta, J. I. and Singh, M. and Welle, T.",
	title = "{Comprehensive simulation of heavy-ion collisions at nonzero baryon chemical potential}",
	eprint = "2206.02655",
	archivePrefix = "arXiv",
	primaryClass = "nucl-th",
	doi = "10.1103/PhysRevC.106.054906",
	journal = "Phys. Rev. C",
	volume = "106",
	number = "5",
	pages = "054906",
	year = "2022"
}

@book{Romatschke:2017ejr,
	author = "Romatschke, Paul and Romatschke, Ulrike",
	title = "{Relativistic Fluid Dynamics In and Out of Equilibrium}",
	eprint = "1712.05815",
	archivePrefix = "arXiv",
	primaryClass = "nucl-th",
	doi = "10.1017/9781108651998",
	isbn = "978-1-108-48368-1, 978-1-108-75002-8",
	publisher = "Cambridge University Press",
	series = "Cambridge Monographs on Mathematical Physics",
	month = "5",
	year = "2019"
}

@article{10.1093/ptep/pts014,
	author = {Nonaka, Chiho and Asakawa, Masayuki},
	title = {Modeling a realistic dynamical model for high energy heavy ion collisions},
	journal = {Progress of Theoretical and Experimental Physics},
	volume = {2012},
	number = {1},
	pages = {01A208},
	year = {2012},
	month = {09},
	issn = {2050-3911},
	doi = {10.1093/ptep/pts014},
	url = {https://doi.org/10.1093/ptep/pts014},
	eprint = {https://academic.oup.com/ptep/article-pdf/2012/1/01A208/4456796/pts014.pdf},
}

@article{Huovinen:2001cy,
    author = "Huovinen, P. and Kolb, P. F. and Heinz, Ulrich W. and Ruuskanen, P. V. and Voloshin, S. A.",
    title = "{Radial and elliptic flow at RHIC: Further predictions}",
    eprint = "hep-ph/0101136",
    archivePrefix = "arXiv",
    doi = "10.1016/S0370-2693(01)00219-2",
    journal = "Phys. Lett. B",
    volume = "503",
    pages = "58--64",
    year = "2001"
}

@article{Heinz:2013th,
    author = "Heinz, Ulrich and Snellings, Raimond",
    title = "{Collective flow and viscosity in relativistic heavy-ion collisions}",
    eprint = "1301.2826",
    archivePrefix = "arXiv",
    primaryClass = "nucl-th",
    doi = "10.1146/annurev-nucl-102212-170540",
    journal = "Ann. Rev. Nucl. Part. Sci.",
    volume = "63",
    pages = "123--151",
    year = "2013"
}

@article{Cleymans:1999st,
    author = "Cleymans, J. and Redlich, K.",
    title = "{Chemical and thermal freezeout parameters from 1-A/GeV to 200-A/GeV}",
    eprint = "nucl-th/9903063",
    archivePrefix = "arXiv",
    reportNumber = "BI-TP-99-19",
    doi = "10.1103/PhysRevC.60.054908",
    journal = "Phys. Rev. C",
    volume = "60",
    pages = "054908",
    year = "1999"
}

@article{Andronic:2005yp,
    author = "Andronic, A. and Braun-Munzinger, P. and Stachel, J.",
    title = "{Hadron production in central nucleus-nucleus collisions at chemical freeze-out}",
    eprint = "nucl-th/0511071",
    archivePrefix = "arXiv",
    doi = "10.1016/j.nuclphysa.2006.03.012",
    journal = "Nucl. Phys. A",
    volume = "772",
    pages = "167--199",
    year = "2006"
}

@article{Cleymans:2005xv,
    author = "Cleymans, J. and Oeschler, H. and Redlich, K. and Wheaton, S.",
    title = "{Comparison of chemical freeze-out criteria in heavy-ion collisions}",
    eprint = "hep-ph/0511094",
    archivePrefix = "arXiv",
    reportNumber = "CERN-PH-TH-2005-210, CERN-PH-TH/2005-210",
    doi = "10.1103/PhysRevC.73.034905",
    journal = "Phys. Rev. C",
    volume = "73",
    pages = "034905",
    year = "2006"
}

@article{Romatschke:2007mq,
	author = "Romatschke, Paul and Romatschke, Ulrike",
	title = "{Viscosity Information from Relativistic Nuclear Collisions: How Perfect is the Fluid Observed at RHIC?}",
	eprint = "0706.1522",
	archivePrefix = "arXiv",
	primaryClass = "nucl-th",
	reportNumber = "INT-PUB-07-14",
	doi = "10.1103/PhysRevLett.99.172301",
	journal = "Phys. Rev. Lett.",
	volume = "99",
	pages = "172301",
	year = "2007"
}

@article{Chen:2024aom,
	author = "Chen, Jinhui and others",
	title = "{Properties of the QCD matter: review of selected results from the relativistic heavy ion collider beam energy scan (RHIC BES) program}",
	eprint = "2407.02935",
	archivePrefix = "arXiv",
	primaryClass = "nucl-ex",
	doi = "10.1007/s41365-024-01591-2",
	journal = "Nucl. Sci. Tech.",
	volume = "35",
	number = "12",
	pages = "214",
	year = "2024"
}

@article{Kolb:2003dz,
	author = "Kolb, Peter F. and Heinz, Ulrich W.",
	editor = "Hwa, Rudolph C. and Wang, Xin-Nian",
	title = "{Hydrodynamic description of ultrarelativistic heavy ion collisions}",
	eprint = "nucl-th/0305084",
	archivePrefix = "arXiv",
	reportNumber = "SUNY-NTG-03-06",
	pages = "634--714",
	month = "5",
	year = "2003"
}

@article{STAR:2017sal,
	author = "Adamczyk, L. and others",
	collaboration = "STAR",
	title = "{Bulk Properties of the Medium Produced in Relativistic Heavy-Ion Collisions from the Beam Energy Scan Program}",
	eprint = "1701.07065",
	archivePrefix = "arXiv",
	primaryClass = "nucl-ex",
	doi = "10.1103/PhysRevC.96.044904",
	journal = "Phys. Rev. C",
	volume = "96",
	number = "4",
	pages = "044904",
	year = "2017"
}

@article{NA49:2002pzu,
	author = "Afanasiev, S. V. and others",
	collaboration = "NA49",
	title = "{Energy dependence of pion and kaon production in central Pb + Pb collisions}",
	eprint = "nucl-ex/0205002",
	archivePrefix = "arXiv",
	doi = "10.1103/PhysRevC.66.054902",
	journal = "Phys. Rev. C",
	volume = "66",
	pages = "054902",
	year = "2002"
}

@article{Cooper:1974mv,
	author = "Cooper, Fred and Frye, Graham",
	title = "{Comment on the Single Particle Distribution in the Hydrodynamic and Statistical Thermodynamic Models of Multiparticle Production}",
	reportNumber = "Print-74-0742 (YESHIVA)",
	doi = "10.1103/PhysRevD.10.186",
	journal = "Phys. Rev. D",
	volume = "10",
	pages = "186",
	year = "1974"
}

@article{Cleymans:1998fq,
	author = "Cleymans, J. and Redlich, K.",
	title = "{Unified description of freezeout parameters in relativistic heavy ion collisions}",
	eprint = "nucl-th/9808030",
	archivePrefix = "arXiv",
	doi = "10.1103/PhysRevLett.81.5284",
	journal = "Phys. Rev. Lett.",
	volume = "81",
	pages = "5284--5286",
	year = "1998"
}

@article{Ouerdane:2002gm,
	author = "Ouerdane, Djamel",
	editor = "Gutbrod, H. and Aichelin, J. and Werner, K.",
	collaboration = "BRAHMS",
	title = "{Rapidity dependence of charged particle yields for Au+Au at $\sqrt{s_{\rm NN}} =$ 200-GeV}",
	eprint = "nucl-ex/0212001",
	archivePrefix = "arXiv",
	doi = "10.1016/S0375-9474(02)01454-9",
	journal = "Nucl. Phys. A",
	volume = "715",
	pages = "478--481",
	year = "2003"
}

@article{HADES:2020ver,
	author = "Adamczewski-Musch, J. and others",
	collaboration = "HADES",
	title = "{Charged-pion production in $\mathbf {Au+Au}$ collisions at $\sqrt{\mathbf {s}_{\mathbf {NN}}} = 2.4~{\mathbf {GeV}}$: HADES Collaboration}",
	eprint = "2005.08774",
	archivePrefix = "arXiv",
	primaryClass = "nucl-ex",
	doi = "10.1140/epja/s10050-020-00237-2",
	journal = "Eur. Phys. J. A",
	volume = "56",
	number = "10",
	pages = "259",
	year = "2020"
}

@article{Bjorken:1982qr,
	author = "Bjorken, J. D.",
	title = "{Highly Relativistic Nucleus-Nucleus Collisions: The Central Rapidity Region}",
	reportNumber = "FERMILAB-PUB-82-044-THY, FERMILAB-PUB-82-044-T",
	doi = "10.1103/PhysRevD.27.140",
	journal = "Phys. Rev. D",
	volume = "27",
	pages = "140--151",
	year = "1983"
}

@article{Ali:2024zvp,
	author = "Ali, Mahammad Sabir and Biswas, Deeptak and Jaiswal, Amaresh and Singh, Sushant K.",
	title = "{Hadron momentum spectra from analytical solutions of relativistic hydrodynamics}",
	eprint = "2403.00624",
	archivePrefix = "arXiv",
	primaryClass = "hep-ph",
	doi = "10.1140/epjc/s10052-025-13751-8",
	journal = "Eur. Phys. J. C",
	volume = "85",
	number = "1",
	pages = "30",
	year = "2025"
}

@article{Florkowski:2004tn,
	author = "Florkowski, Wojciech and Broniowski, Wojciech",
	editor = "Sadzikowski, M.",
	title = "{Hydro-inspired parameterizations of freeze-out in relativistic heavy-ion collisions}",
	eprint = "nucl-th/0410081",
	archivePrefix = "arXiv",
	journal = "Acta Phys. Polon. B",
	volume = "35",
	pages = "2895--2910",
	year = "2004"
}

@article{Schnedermann:1993ws,
	author = "Schnedermann, Ekkard and Sollfrank, Josef and Heinz, Ulrich W.",
	title = "{Thermal phenomenology of hadrons from 200-A/GeV S+S collisions}",
	eprint = "nucl-th/9307020",
	archivePrefix = "arXiv",
	reportNumber = "TPR-93-16",
	doi = "10.1103/PhysRevC.48.2462",
	journal = "Phys. Rev. C",
	volume = "48",
	pages = "2462--2475",
	year = "1993"
}

@article{Melo:2019mpn,
	author = "Melo, Ivan and Tom{\'a}{\v{s}}ik, Boris",
	title = "{Kinetic freeze-out in central heavy-ion collisions between 7.7 and 2760 GeV per nucleon pair}",
	eprint = "1908.03023",
	archivePrefix = "arXiv",
	primaryClass = "nucl-th",
	doi = "10.1088/1361-6471/ab5f03",
	journal = "J. Phys. G",
	volume = "47",
	number = "4",
	pages = "045107",
	year = "2020"
}

@article{STAR:2001ksn,
	author = "Adler, C. and others",
	collaboration = "STAR",
	title = "{Identified particle elliptic flow in Au + Au collisions at $\sqrt{s_{\rm NN}} = 130$-GeV}",
	eprint = "nucl-ex/0107003",
	archivePrefix = "arXiv",
	doi = "10.1103/PhysRevLett.87.182301",
	journal = "Phys. Rev. Lett.",
	volume = "87",
	pages = "182301",
	year = "2001"
}

@article{Retiere:2003kf,
	author = "Retiere, Fabrice and Lisa, Michael Annan",
	title = "{Observable implications of geometrical and dynamical aspects of freeze out in heavy ion collisions}",
	eprint = "nucl-th/0312024",
	archivePrefix = "arXiv",
	doi = "10.1103/PhysRevC.70.044907",
	journal = "Phys. Rev. C",
	volume = "70",
	pages = "044907",
	year = "2004"
}

@article{He:2010vw,
	author = "He, Min and Fries, Rainer J. and Rapp, Ralf",
	title = "{Scaling of Elliptic Flow, Recombination and Sequential Freeze-Out of Hadrons in Heavy-Ion Collisions}",
	eprint = "1006.1111",
	archivePrefix = "arXiv",
	primaryClass = "nucl-th",
	doi = "10.1103/PhysRevC.82.034907",
	journal = "Phys. Rev. C",
	volume = "82",
	pages = "034907",
	year = "2010"
}

@article{Sun:2014rda,
	author = "Sun, X. and Masui, H. and Poskanzer, A. M. and Schmah, A.",
	title = "{Blast Wave Fits to Elliptic Flow Data at $\sqrt{s_{\rm NN}} =$ 7.7--2760 GeV}",
	eprint = "1410.1947",
	archivePrefix = "arXiv",
	primaryClass = "hep-ph",
	doi = "10.1103/PhysRevC.91.024903",
	journal = "Phys. Rev. C",
	volume = "91",
	number = "2",
	pages = "024903",
	year = "2015"
}

@article{Cimerman:2017lmm,
	author = "Cimerman, Jakub and Tomasik, Boris and Csanad, Mate and Lokos, Sandor",
	title = "{Higher-order anisotropies in the Blast-Wave Model - disentangling flow and density field anisotropies}",
	eprint = "1702.01735",
	archivePrefix = "arXiv",
	primaryClass = "nucl-th",
	doi = "10.1140/epja/i2017-12349-7",
	journal = "Eur. Phys. J. A",
	volume = "53",
	number = "8",
	pages = "161",
	year = "2017"
}

@article{Melo:2015wpa,
	author = "Melo, Ivan and Tomasik, Boris",
	title = "{Reconstructing the final state of Pb+Pb collisions at $\sqrt{s_{NN}}=2.76$ TeV}",
	eprint = "1502.01247",
	archivePrefix = "arXiv",
	primaryClass = "nucl-th",
	doi = "10.1088/0954-3899/43/1/015102",
	journal = "J. Phys. G",
	volume = "43",
	number = "1",
	pages = "015102",
	year = "2016"
}

@article{Tomasik:2024uuq,
	author = "Tomasik, Boris",
	title = "{On elliptic flow and the blast-wave model}",
	eprint = "2409.19758",
	archivePrefix = "arXiv",
	primaryClass = "nucl-th",
	doi = "10.1142/S0217751X25420060",
	journal = "Int. J. Mod. Phys. A",
	volume = "40",
	number = "21",
	pages = "2542006",
	year = "2025"
}

@article{FLORKOWSKI2025100249,
	title = {Statistical hadronization model for low-energy heavy-ion collisions},
	journal = {Journal of Subatomic Particles and Cosmology},
	volume = {4},
	pages = {100249},
	year = {2025},
	issn = {3050-4805},
	doi = {https://doi.org/10.1016/j.jspc.2025.100249},
	url = {https://www.sciencedirect.com/science/article/pii/S3050480525002298},
	author = {Wojciech Florkowski and Radoslaw Ryblewski}
}

@article{Broniowski:2001we,
	author = "Broniowski, Wojciech and Florkowski, Wojciech",
	title = "{Explanation of the RHIC p(T) spectra in a thermal model with expansion}",
	eprint = "nucl-th/0106050",
	archivePrefix = "arXiv",
	doi = "10.1103/PhysRevLett.87.272302",
	journal = "Phys. Rev. Lett.",
	volume = "87",
	pages = "272302",
	year = "2001"
}

@article{PhysRevLett.42.880,
	title = {Evidence for a Blast Wave from Compressed Nuclear Matter},
	author = {Siemens, Philip J. and Rasmussen, John O.},
	journal = {Phys. Rev. Lett.},
	volume = {42},
	issue = {14},
	pages = {880--883},
	numpages = {0},
	year = {1979},
	month = {Apr},
	publisher = {American Physical Society},
	doi = {10.1103/PhysRevLett.42.880},
	url = {https://link.aps.org/doi/10.1103/PhysRevLett.42.880}
}

@article{Harabasz:2020sei,
	author = "Harabasz, Szymon and Florkowski, Wojciech and Galatyuk, Tetyana and Ma Lgorzata Gumberidze, {\textdaggerdbl}. and Ryblewski, Radoslaw and Salabura, Piotr and Stroth, Joachim",
	title = "{Statistical hadronization model for heavy-ion collisions in the few-GeV energy regime}",
	eprint = "2003.12992",
	archivePrefix = "arXiv",
	primaryClass = "nucl-th",
	doi = "10.1103/PhysRevC.102.054903",
	journal = "Phys. Rev. C",
	volume = "102",
	number = "5",
	pages = "054903",
	year = "2020"
}

@article{Harabasz:2022rdt,
	author = "Harabasz, Szymon and Ko{\l}a{\'s}, Jedrzej and Ryblewski, Rados{\l}aw and Florkowski, Wojciech and Galatyuk, Tetyana and Gumberidze, Ma{\l}gorzata and Salabura, Piotr and Stroth, Joachim and Zbroszczyk, Hanna Paulina",
	title = "{Spheroidal expansion and freeze-out geometry of heavy-ion collisions in the few-GeV energy regime}",
	eprint = "2210.07694",
	archivePrefix = "arXiv",
	primaryClass = "nucl-th",
	doi = "10.1103/PhysRevC.107.034917",
	journal = "Phys. Rev. C",
	volume = "107",
	number = "3",
	pages = "034917",
	year = "2023"
}

@article{Broniowski:2001uk,
	author = "Broniowski, Wojciech and Florkowski, Wojciech",
	title = "{Strange particle production at RHIC in a single freezeout model}",
	eprint = "nucl-th/0112043",
	archivePrefix = "arXiv",
	doi = "10.1103/PhysRevC.65.064905",
	journal = "Phys. Rev. C",
	volume = "65",
	pages = "064905",
	year = "2002"
}

@article{Drogosz:2025vdq,
	author = "Drogosz, Zbigniew and Florkowski, Wojciech and Witkowski, Nikodem and Ryblewski, Radoslaw",
	title = "{$^{3}$H and $^{3}$He nuclei production in a combined thermal and coalescence framework for heavy-ion collisions in the few-GeV energy regime*}",
	eprint = "2504.00283",
	archivePrefix = "arXiv",
	primaryClass = "hep-ph",
	doi = "10.1088/1674-1137/ae099a",
	journal = "Chin. Phys.",
	volume = "50",
	number = "1",
	pages = "014104",
	year = "2026"
}

@article{Teaney:2003kp,
	author = "Teaney, Derek",
	title = "{The Effects of viscosity on spectra, elliptic flow, and HBT radii}",
	eprint = "nucl-th/0301099",
	archivePrefix = "arXiv",
	doi = "10.1103/PhysRevC.68.034913",
	journal = "Phys. Rev. C",
	volume = "68",
	pages = "034913",
	year = "2003"
}

@article{Jaiswal:2015saa,
	author = "Jaiswal, Amaresh and Koch, Volker",
	title = "{A viscous blast-wave model for relativistic heavy-ion collisions}",
	eprint = "1508.05878",
	archivePrefix = "arXiv",
	primaryClass = "nucl-th",
	month = "8",
	year = "2015"
}

@article{Yang:2016rnw,
	author = "Yang, Z. and Fries, Rainer J.",
	title = "{A Blast Wave Model With Viscous Corrections}",
	eprint = "1612.05629",
	archivePrefix = "arXiv",
	primaryClass = "nucl-th",
	doi = "10.1088/1742-6596/832/1/012056",
	journal = "J. Phys. Conf. Ser.",
	volume = "832",
	number = "1",
	pages = "012056",
	year = "2017"
}

@article{Yang:2023apw,
	author = "Yang, Zhidong and Sun, Yifeng and Chen, Lie-Wen",
	title = "{Baryon-density dependence of viscosities of the quark-gluon plasma at hadronization}",
	eprint = "2310.17444",
	archivePrefix = "arXiv",
	primaryClass = "nucl-th",
	doi = "10.1103/PhysRevC.109.054907",
	journal = "Phys. Rev. C",
	volume = "109",
	number = "5",
	pages = "054907",
	year = "2024"
}

@article{Yang:2022ixy,
	author = "Yang, Zhidong and Chen, Lie-Wen",
	title = "{Bayesian inference of the specific shear and bulk viscosities of the quark-gluon plasma at crossover from {\ensuremath{\phi}} and {\ensuremath{\Omega}} observables}",
	eprint = "2207.13534",
	archivePrefix = "arXiv",
	primaryClass = "nucl-th",
	doi = "10.1103/PhysRevC.107.064910",
	journal = "Phys. Rev. C",
	volume = "107",
	number = "6",
	pages = "064910",
	year = "2023"
}

@article{Yang:2018ghi,
title = {Shear stress tensor and specific shear viscosity of hot hadron gas in nuclear collisions},
author = {Yang, Zhidong and Fries, Rainer J.},
journal = {Phys. Rev. C},
volume = {105},
issue = {1},
pages = {014910},
numpages = {14},
year = {2022},
month = {Jan},
publisher = {American Physical Society},
doi = {10.1103/PhysRevC.105.014910},
url = {https://link.aps.org/doi/10.1103/PhysRevC.105.014910}
}

@article{Yang:2020oig,
	author = "Yang, Zhidong and Fries, Rainer J.",
	title = "{Parameterizing smooth viscous fluid dynamics with a viscous blast wave}",
	eprint = "2007.11777",
	archivePrefix = "arXiv",
	primaryClass = "nucl-th",
	doi = "10.1088/1361-6471/ad0914",
	journal = "J. Phys. G",
	volume = "51",
	number = "1",
	pages = "015102",
	year = "2024"
}

@article{Tang:2008ud,
	author = "Tang, Zebo and Xu, Yichun and Ruan, Lijuan and van Buren, Gene and Wang, Fuqiang and Xu, Zhangbu",
	title = "{Spectra and radial flow at RHIC with Tsallis statistics in a Blast-Wave description}",
	eprint = "0812.1609",
	archivePrefix = "arXiv",
	primaryClass = "nucl-ex",
	reportNumber = "BNL-KB-02-02",
	doi = "10.1103/PhysRevC.79.051901",
	journal = "Phys. Rev. C",
	volume = "79",
	pages = "051901",
	year = "2009"
}

@article{Rode:2018hlj,
	author = "Rode, Sudhir Pandurang and Bhaduri, Partha Pratim and Jaiswal, Amaresh and Roy, Ankhi",
	title = "{Kinetic freeze-out conditions in nuclear collisions with $2A$ - $158A$ GeV beam energy within a non-boost-invariant blast-wave model}",
	eprint = "1805.11463",
	archivePrefix = "arXiv",
	primaryClass = "nucl-th",
	doi = "10.1103/PhysRevC.98.024907",
	journal = "Phys. Rev. C",
	volume = "98",
	number = "2",
	pages = "024907",
	year = "2018"
}

@article{Rode:2020vhu,
	author = "Rode, Sudhir Pandurang and Bhaduri, Partha Pratim and Jaiswal, Amaresh and Roy, Ankhi",
	title = "{Hierarchy of kinetic freeze-out parameters in low energy heavy-ion collisions}",
	eprint = "2004.04703",
	archivePrefix = "arXiv",
	primaryClass = "hep-ph",
	doi = "10.1103/PhysRevC.102.054912",
	journal = "Phys. Rev. C",
	volume = "102",
	number = "5",
	pages = "054912",
	year = "2020"
}

@article{Chatterjee:2014lfa,
	author = "Chatterjee, Sandeep and Mohanty, Bedangadas and Singh, Ranbir",
	title = "{Freezeout hypersurface at energies available at the CERN Large Hadron Collider from particle spectra: Flavor and centrality dependence}",
	eprint = "1411.1718",
	archivePrefix = "arXiv",
	primaryClass = "nucl-th",
	doi = "10.1103/PhysRevC.92.024917",
	journal = "Phys. Rev. C",
	volume = "92",
	number = "2",
	pages = "024917",
	year = "2015"
}

@article{Alam:2026ixb,
	author = "Alam, Sk Noor and Roy, Victor",
	title = "{Kinetic Freeze-Out Conditions and Net Baryon Density in Au+Au Collisions at $\sqrt{s_{NN}} = 7.7$--$39$ GeV within a Collective Flow Fireball Model}",
	eprint = "2603.07160",
	archivePrefix = "arXiv",
	primaryClass = "nucl-th",
	month = "3",
	year = "2026"
}

@article{Parvan:2026idk,
	author = "Parvan, A. S. and Aparin, A. A. and Nedorezov, E. V.",
	title = "{Finite Volume Effects on Transverse Momentum Spectra at LHC and RHIC Using a Blast-Wave Model with Planck Transformed Temperatures}",
	eprint = "2604.07410",
	archivePrefix = "arXiv",
	primaryClass = "hep-ph",
	month = "4",
	year = "2026"
}

@article{Rai:2026dda,
	author = "Rai, Anand and Dwibedi, Ashutosh and Ghosh, Sabyasachi",
	title = "{Spectra and elliptic flow of light hadrons in an expanding fire-cylinder model for the RHIC Beam Energy Scan}",
	eprint = "2602.17241",
	archivePrefix = "arXiv",
	primaryClass = "nucl-th",
	doi = "10.1016/j.nuclphysa.2026.123440",
	journal = "Nucl. Phys. A",
	volume = "1073",
	pages = "123440",
	year = "2026"
}

@article{Prasad:2021bdq,
	author = "Prasad, Suraj and Mallick, Neelkamal and Behera, Debadatta and Sahoo, Raghunath and Tripathy, Sushanta",
	title = "{Event topology and global observables in heavy-ion collisions at the Large Hadron Collider}",
	eprint = "2112.03892",
	archivePrefix = "arXiv",
	primaryClass = "hep-ph",
	doi = "10.1038/s41598-022-07547-z",
	journal = "Sci. Rep.",
	volume = "12",
	number = "1",
	pages = "3917",
	year = "2022"
}

@article{Panda:2025lmd,
	author = "Panda, Ankit Kumar",
	title = "{Dominance of electric fields in the charge splitting of elliptic flow}",
	eprint = "2501.07240",
	archivePrefix = "arXiv",
	primaryClass = "hep-ph",
	doi = "10.1088/1361-6471/adce1b",
	journal = "J. Phys. G",
	volume = "52",
	number = "5",
	pages = "055102",
	year = "2025"
}

@article{Rapp:1999us,
	author = "Rapp, Ralf and Wambach, Jochen",
	title = "{Low mass dileptons at the CERN SPS: Evidence for chiral restoration?}",
	eprint = "hep-ph/9907502",
	archivePrefix = "arXiv",
	reportNumber = "SUNY-NTG-99-25",
	doi = "10.1007/s100500050364",
	journal = "Eur. Phys. J. A",
	volume = "6",
	pages = "415--420",
	year = "1999"
}

@article{Rapp:1999zw,
	author = "Rapp, Ralf and Shuryak, Edward V.",
	title = "{Thermal dilepton radiation at intermediate masses at the CERN - SPS}",
	eprint = "hep-ph/9909348",
	archivePrefix = "arXiv",
	reportNumber = "SUNY-NTG-99-30",
	doi = "10.1016/S0370-2693(99)01367-2",
	journal = "Phys. Lett. B",
	volume = "473",
	pages = "13--19",
	year = "2000"
}

@article{Rapp:2000pe,
	author = "Rapp, R.",
	title = "{Signatures of thermal dilepton radiation at RHIC}",
	eprint = "hep-ph/0010101",
	archivePrefix = "arXiv",
	reportNumber = "SUNY-NTG-00-41",
	doi = "10.1103/PhysRevC.63.054907",
	journal = "Phys. Rev. C",
	volume = "63",
	pages = "054907",
	year = "2001"
}

@article{Turbide:2003si,
	author = "Turbide, Simon and Rapp, Ralf and Gale, Charles",
	title = "{Hadronic production of thermal photons}",
	eprint = "hep-ph/0308085",
	archivePrefix = "arXiv",
	reportNumber = "NORDITA-2003-51-NP",
	doi = "10.1103/PhysRevC.69.014903",
	journal = "Phys. Rev. C",
	volume = "69",
	pages = "014903",
	year = "2004"
}

@article{vanHees:2005wb,
	author = "van Hees, Hendrik and Greco, Vincenzo and Rapp, Ralf",
	title = "{Heavy-quark probes of the quark-gluon plasma and interpretation of recent data taken at the BNL Relativistic Heavy Ion Collider}",
	eprint = "nucl-th/0508055",
	archivePrefix = "arXiv",
	doi = "10.1103/PhysRevC.73.034913",
	journal = "Phys. Rev. C",
	volume = "73",
	pages = "034913",
	year = "2006"
}

@article{vanHees:2006ng,
	author = "van Hees, Hendrik and Rapp, Ralf",
	title = "{Comprehensive interpretation of thermal dileptons at the SPS}",
	eprint = "hep-ph/0603084",
	archivePrefix = "arXiv",
	doi = "10.1103/PhysRevLett.97.102301",
	journal = "Phys. Rev. Lett.",
	volume = "97",
	pages = "102301",
	year = "2006"
}

@article{vanHees:2007th,
	author = "van Hees, Hendrik and Rapp, Ralf",
	title = "{Dilepton Radiation at the CERN Super Proton Synchrotron}",
	eprint = "0711.3444",
	archivePrefix = "arXiv",
	primaryClass = "hep-ph",
	doi = "10.1016/j.nuclphysa.2008.03.009",
	journal = "Nucl. Phys. A",
	volume = "806",
	pages = "339--387",
	year = "2008"
}

@article{vanHees:2011vb,
	author = "van Hees, Hendrik and Gale, Charles and Rapp, Ralf",
	title = "{Thermal Photons and Collective Flow at the Relativistic Heavy-Ion Collider}",
	eprint = "1108.2131",
	archivePrefix = "arXiv",
	primaryClass = "hep-ph",
	doi = "10.1103/PhysRevC.84.054906",
	journal = "Phys. Rev. C",
	volume = "84",
	pages = "054906",
	year = "2011"
}

@article{Gossiaux:2011ea,
	author = "Gossiaux, Pol Bernard and Vogel, Sascha and van Hees, Hendrik and Aichelin, Joerg and Rapp, Ralf and He, Min and Bluhm, Marcus",
	title = "{The Influence of bulk evolution models on heavy-quark phenomenology}",
	eprint = "1102.1114",
	archivePrefix = "arXiv",
	primaryClass = "hep-ph",
	month = "2",
	year = "2011"
}

@article{vanHees:2014ida,
	author = "van Hees, Hendrik and He, Min and Rapp, Ralf",
	title = "{Pseudo-critical enhancement of thermal photons in relativistic heavy-ion collisions?}",
	eprint = "1404.2846",
	archivePrefix = "arXiv",
	primaryClass = "nucl-th",
	doi = "10.1016/j.nuclphysa.2014.09.009",
	journal = "Nucl. Phys. A",
	volume = "933",
	pages = "256--271",
	year = "2015"
}

@article{Rapp:2014hha,
	author = "Rapp, Ralf and van Hees, Hendrik",
	title = "{Thermal Dileptons as Fireball Thermometer and Chronometer}",
	eprint = "1411.4612",
	archivePrefix = "arXiv",
	primaryClass = "hep-ph",
	doi = "10.1016/j.physletb.2015.12.065",
	journal = "Phys. Lett. B",
	volume = "753",
	pages = "586--590",
	year = "2016"
}

@article{PhysRevLett.93.102301,
	title = {Nuclear Stopping in $\mathrm{A}\mathrm{u}+\mathrm{A}\mathrm{u}$ Collisions at $\sqrt{{s}_{NN}}=200\text{ }\text{ }\mathrm{G}\mathrm{e}\mathrm{V}$},
	author = {Bearden, I. G. and Beavis, D. and Besliu, C. and Budick, B. and B\o{}ggild, H. and Chasman, C. and Christensen, C. H. and Christiansen, P. and Cibor, J. and Debbe, R. and Enger, E. and Gaardh\o{}je, J. J. and Germinario, M. and Hagel, K. and Hansen, O. and Holm, A. and Holme, A. K. and Ito, H. and Jipa, A. and Jundt, F. and J\o{}rdre, J. I. and J\o{}rgensen, C. E. and Karabowicz, R. and Kim, E. J. and Kozik, T. and Larsen, T. M. and Lee, J. H. and Lee, Y. K. and L\o{}vh\o{}iden, G. and Majka, Z. and Makeev, A. and Mikelsen, M. and Murray, M. and Natowitz, J. and Nielsen, B. S. and Norris, J. and Olchanski, K. and Ouerdane, D. and P\l{}aneta, R. and Rami, F. and Ristea, C. and R\"ohrich, D. and Samset, B. H. and Sandberg, D. and Sanders, S. J. and Scheetz, R. A. and Staszel, P. and Tveter, T. S. and Videb\ae{}k, F. and Wada, R. and Yin, Z. and Zgura, I. S.},
	collaboration = {BRAHMS Collaboration},
	journal = {Phys. Rev. Lett.},
	volume = {93},
	issue = {10},
	pages = {102301},
	numpages = {5},
	year = {2004},
	month = {Aug},
	publisher = {American Physical Society},
	doi = {10.1103/PhysRevLett.93.102301},
	url = {https://link.aps.org/doi/10.1103/PhysRevLett.93.102301}
}

@article{PhysRevLett.94.162301,
	title = {Charged Meson Rapidity Distributions in Central $\mathrm{Au}+\mathrm{Au}$ Collisions at $\sqrt{{s}_{NN}}=200\text{ }\text{ }\mathrm{GeV}$},
	author = {Bearden, I. G. and Beavis, D. and Besliu, C. and Budick, B. and B\o{}ggild, H. and Chasman, C. and Christensen, C. H. and Christiansen, P. and Cibor, J. and Debbe, R. and Enger, E. and Gaardh\o{}je, J. J. and Germinario, M. and Hagel, K. and Hansen, O. and Holm, A. and Holme, A. K. and Ito, H. and Jipa, A. and Jundt, F. and J\o{}rdre, J. I. and J\o{}rgensen, C. E. and Karabowicz, R. and Kim, E. J. and Kozik, T. and Larsen, T. M. and Lee, J. H. and Lee, Y. K. and L\o{}vh\o{}iden, G. and Majka, Z. and Makeev, A. and Mikelsen, M. and Murray, M. and Natowitz, J. and Nielsen, B. S. and Norris, J. and Olchanski, K. and Ouerdane, D. and P\l{}aneta, R. and Rami, F. and Ristea, C. and R\"ohrich, D. and Samset, B. H. and Sandberg, D. and Sanders, S. J. and Sheetz, R. A. and Staszel, P. and Tveter, T. S. and Videb\ae{}k, F. and Wada, R. and Yin, Z. and Zgura, I. S.},
	collaboration = {BRAHMS Collaboration},
	journal = {Phys. Rev. Lett.},
	volume = {94},
	issue = {16},
	pages = {162301},
	numpages = {4},
	year = {2005},
	month = {Apr},
	publisher = {American Physical Society},
	doi = {10.1103/PhysRevLett.94.162301},
	url = {https://link.aps.org/doi/10.1103/PhysRevLett.94.162301}
}

@article{Bondorf:1978kz,
	author = "Bondorf, J. P. and Garpman, S. I. A. and Zimanyi, J.",
	title = "{A Simple Analytic Hydrodynamic Model for Expanding Fireballs}",
	doi = "10.1016/0375-9474(78)90076-3",
	journal = "Nucl. Phys. A",
	volume = "296",
	pages = "320--332",
	year = "1978"
}

@article{Wheaton_2009,
	title={THERMUS—A thermal model package for ROOT},
	volume={180},
	ISSN={0010-4655},
	url={http://dx.doi.org/10.1016/j.cpc.2008.08.001},
	DOI={10.1016/j.cpc.2008.08.001},
	number={1},
	journal={Computer Physics Communications},
	publisher={Elsevier BV},
	author={Wheaton, S. and Cleymans, J. and Hauer, M.},
	year={2009},
	month=Jan, pages={84–106} }

@article{Becattini:2000jw,
	author = "Becattini, F. and Cleymans, J. and Keranen, A. and Suhonen, E and Redlich, K.",
	title = "{Features of particle multiplicities and strangeness production in central heavy ion collisions between 1.7A-GeV/c and 158A-GeV/c}",
	eprint = "hep-ph/0002267",
	archivePrefix = "arXiv",
	reportNumber = "BI-TP-00-02, DFF-349-02-2000",
	doi = "10.1103/PhysRevC.64.024901",
	journal = "Phys. Rev. C",
	volume = "64",
	pages = "024901",
	year = "2001"
}

@article{Braun-Munzinger:2003pwq,
	author = "Braun-Munzinger, Peter and Redlich, Krzysztof and Stachel, Johanna",
	editor = "Hwa, Rudolph C. and Wang, Xin-Nian",
	title = "{Particle production in heavy ion collisions}",
	eprint = "nucl-th/0304013",
	archivePrefix = "arXiv",
	reportNumber = "GSI-PREPRINT-2003-13",
	doi = "10.1142/9789812795533_0008",
	pages = "491--599",
	month = "4",
	year = "2003"
}

@article{Chatterjee:2015fua,
	author = "Chatterjee, Sandeep and Das, Sabita and Kumar, Lokesh and Mishra, D. and Mohanty, Bedangadas and Sahoo, Raghunath and Sharma, Natasha",
	title = "{Freeze-Out Parameters in Heavy-Ion Collisions at AGS, SPS, RHIC, and LHC Energies}",
	doi = "10.1155/2015/349013",
	journal = "Adv. High Energy Phys.",
	volume = "2015",
	pages = "349013",
	year = "2015"
}

@article{Chatterjee:2017yhp,
	author = "Chatterjee, Sandeep and Mishra, Debadeepti and Mohanty, Bedangadas and Samanta, Subhasis",
	title = "{Freezeout systematics due to the hadron spectrum}",
	eprint = "1708.08152",
	archivePrefix = "arXiv",
	primaryClass = "nucl-th",
	doi = "10.1103/PhysRevC.96.054907",
	journal = "Phys. Rev. C",
	volume = "96",
	number = "5",
	pages = "054907",
	year = "2017"
}

@article{STAR:2008med,
	author = "Abelev, B. I. and others",
	collaboration = "STAR",
	title = "{Systematic Measurements of Identified Particle Spectra in $p p, d^+$ Au and Au+Au Collisions from STAR}",
	eprint = "0808.2041",
	archivePrefix = "arXiv",
	primaryClass = "nucl-ex",
	doi = "10.1103/PhysRevC.79.034909",
	journal = "Phys. Rev. C",
	volume = "79",
	pages = "034909",
	year = "2009"
}
\end{document}